\documentclass[dvips]{acta}
\usepackage{supertabular,lscape,epsfig}
\usepackage{amssymb}
\usepackage{amsmath}
\DeclareSymbolFont{ppa}{OT1}{ppl}{m}{it}
\DeclareMathSymbol{\vv}{\mathalpha}{ppa}{'166}

\newfont{\hb}{rphvb at 10pt}%bezszeryfowe pó³grube
\newfont{\hbo}{rphvbo at 10pt}%bezszeryfowe pó³grube kursywa
\newfont{\bitt}{rptmbi at 12pt}%pó³gruba kursywa (tytu³ artyku³u)
\newfont{\bits}{rptmbi at 11pt}%pó³gruba kursywa (tytu³y rozdzia³ów)

\SetPages{1}{21}

\SetVol{63}{2013}

\begin{document}

%Zwarte naglowki, jeden wiersz, appendix
\newcommand{\TabApp}[2]{\begin{center}\parbox[t]{#1}{\centerline{
  {\bf Appendix}}
  \vskip2mm
  \centerline{\small {\spaceskip 2pt plus 1pt minus 1pt T a b l e}
  \refstepcounter{table}\thetable}
  \vskip2mm
  \centerline{\footnotesize #2}}
  \vskip3mm
\end{center}}

%Zwarte naglowki, jeden wiersz
\newcommand{\TabCapp}[2]{\begin{center}\parbox[t]{#1}{\centerline{
  \small {\spaceskip 2pt plus 1pt minus 1pt T a b l e}
  \refstepcounter{table}\thetable}
  \vskip2mm
  \centerline{\footnotesize #2}}
  \vskip3mm
\end{center}}

%Zwarte naglowki, dwa wiersze
\newcommand{\TTabCap}[3]{\begin{center}\parbox[t]{#1}{\centerline{
  \small {\spaceskip 2pt plus 1pt minus 1pt T a b l e}
  \refstepcounter{table}\thetable}
  \vskip2mm
  \centerline{\footnotesize #2}
  \centerline{\footnotesize #3}}
  \vskip1mm
\end{center}}

%Zwarte naglowki, jeden wiersz, appendix
\newcommand{\MakeTableApp}[4]{\begin{table}[p]\TabApp{#2}{#3}
  \begin{center} \TableFont \begin{tabular}{#1} #4 
  \end{tabular}\end{center}\end{table}}

%Zwarte naglowki, jeden wiersz
\newcommand{\MakeTableSepp}[4]{\begin{table}[p]\TabCapp{#2}{#3}
  \begin{center} \TableFont \begin{tabular}{#1} #4 
  \end{tabular}\end{center}\end{table}}

%Zwarte naglowki, jeden wiersz
\newcommand{\MakeTableee}[4]{\begin{table}[htb]\TabCapp{#2}{#3}
  \begin{center} \TableFont \begin{tabular}{#1} #4
  \end{tabular}\end{center}\end{table}}

%Zwarte naglowki, dwa wiersze
\newcommand{\MakeTablee}[5]{\begin{table}[htb]\TTabCap{#2}{#3}{#4}
  \begin{center} \TableFont \begin{tabular}{#1} #5 
  \end{tabular}\end{center}\end{table}}

%{\it Acta Astronomica Archive}
%\parskip=0pt \itemsep=1mm \setlength{\itemsep}{0.4mm}\setlength{\parindent}{-1em} \setlength{\itemindent}{-1em} - po \begin{itemize} - wszystko
%FWHM, PSF, S/N - proste, 
%MgII, H$\alpha$
%rms, rhs, sd - kursywa
%{\sc DAOPhot}
%{\sc Fnpeaks}
%{\sf files}
%Galactic wszystko (bulge, center, plane, disk, coordinates, latitudes...)
%Cepheids
%type~ Cepheids, Population~II Cepheids
%a.u.
\newfont{\bb}{ptmbi8t at 12pt}
\newfont{\bbb}{cmbxti10}
\newfont{\bbbb}{cmbxti10 at 9pt}
\newcommand{\uprule}{\rule{0pt}{2.5ex}}
\newcommand{\douprule}{\rule[-2ex]{0pt}{4.5ex}}
\newcommand{\dorule}{\rule[-2ex]{0pt}{2ex}}
\def\thefootnote{\fnsymbol{footnote}}
\begin{Titlepage}
\Title{The Variable Stars from the OGLE-III Shallow Survey\\
in the Large Magellanic Cloud\footnote{Based on observations obtained with
the 1.3 m Warsaw telescope at the Las Campanas Observatory of the Carnegie
Institution for Science.}}
\vspace*{7pt}
\Author{K.~~U~l~a~c~z~y~k$^1$,~~ M.\,K.~~S~z~y~m~a~ñ~s~k~i$^1$,~~ A.~~U~d~a~l~s~k~i$^1$,\\
M.~~K~u~b~i~a~k$^1$, ~~G.~~P~i~e~t~r~z~y~ñ~s~k~i$^{1,2}$, ~~ I.~~S~o~s~z~y~ñ~s~k~i$^1$,
£.~~W~y~r~z~y~k~o~w~s~k~i$^{1,3}$, ~~ R.~~P~o~l~e~s~k~i$^{1,4}$,\\
W.~~G~i~e~r~e~n$^2$,~~ A.\,R.~~W~a~l~k~e~r$^5$ ~~and~~ A.~~G~a~r~c~i~a~-~V~a~r~e~l~a$^6$}
{$^1$Warsaw University Observatory, Al.~Ujazdowskie~4, 00-478~Warszawa, Poland\\
e-mail: (kulaczyk,msz,udalski,mk,pietrzyn,soszynsk,wyrzykow,rpoleski)@astrouw.edu.pl\\
$^2$Universidad de Concepción, Departamento de Astronomia, Casilla 160-C, Concepción, Chile\\
e-mail: wgieren@astro-udec.cl\\
$^3$Institute of Astronomy, University of Cambridge, Madingley Road, Cambridge CB3~0HA, UK\\
$^4$Department of Astronomy, The Ohio State University, 140 W. 18th Ave., Columbus, OH 43210, USA\\
$^5$Cerro Tololo Inter-American Observatory, Casilla~603, La~Serena, Chile\\
e-mail: awalker@ctio.noao.edu\\
$^6$Departamento de Fisica, Universidad de los Andes, Bogota, Colombia\\
e-mail: josegarc@uniandes.edu.co} 
\Received{March 21, 2013}
\end{Titlepage}

\vspace*{-10pt}

\Abstract{We describe variable stars found in the data collected
during the OGLE-III Shallow Survey covering the {\it I}-band magnitude
range from 9.7~mag to 14.5~mag. The main result is the extension of
period--luminosity relations for Cepheids up to 134 days. We also detected
82 binary systems and 110 long-period variables not present in the main
OGLE catalogs. Additionally 558 objects were selected as candidates for
miscellaneous variables.}{Magellanic Clouds -- Surveys -- Catalogs --
Cepheids -- Stars: late type -- binaries: eclipsing variables}

\vspace*{-9pt}
\Section{Introduction}
\vspace*{-5pt}
In the previous paper (Ulaczyk \etal 2012) we described reduction
procedures applied to the data obtained in the course of the OGLE-III
Shallow Survey in the Large Magellanic Cloud. We have also published
photometric maps and discussed the quality of photometric data.

Here we present results of variability search based on the same survey. So
far most numerous and extensive catalogs of variable stars for that galaxy
were published based on the OGLE-III data (\eg Soszyñski \etal 2008, 2009a,
2009b). In order to improve even further their completeness we decided to
include bright variable objects of luminosities up to the Shallow Survey
saturation {\it I}-magnitude of 9.3. The main purpose was to supplement the
OGLE-III catalogs with bright Cepheids to derive consistent
luminosity--period relations for wider range of periods. The survey was
also focused on red giant variables and luminous eclipsing stars. In the
following sections we describe subsequent types of identified variable
stars.

\Section{Observational Data}
\vspace*{-2pt}
As it was described in the previous paper photometric data were collected
using the 1.3-m Warsaw Telescope located at Las Campanas Observatory,
operated by the Carnegie Institution for Science. The mosaic camera
consisted of eight CCD chips covering approximately $35\arcm\times35\arcm$
field of view with the scale of 0.26~arcsec/pixel. Observations were evenly
carried out through the {\it I} and {\it V} filters closely resembling
standard filters, although the {\it I}-band filter had higher transmission
for longer wavelengths, which was appropriately corrected during the
reductions based on calibrated OGLE-IV photometry (Szymañski \etal
2011). We used exactly the same photometric system as in the OGLE-III main
survey.  Detailed information about whole instrumentation can be found in
Udalski (2003).

Images were collected during the nights of seeing worse than about
2\arcs. It allowed to blur profiles of the brightest stars preventing them
from saturation. Moreover, we were able to perform observations during the
nights of poor weather which normally would be lost for the main OGLE
survey. Details on reduction procedures were described in the previous
paper (Ulaczyk \etal 2012).

\Section{Variability Search}
\vspace*{-2pt}

Using {\sc Fnpeaks} program created by Z.~Ko³aczkowski we calculated power
spectrum of periods for 65\,989 stars having over 10 observational points
and with average {\it I}-band magnitude brighter than 14.5~mag. The
frequency range from 0.002 to 2 cycles per day was tested with a step of
0.00001.  For every star highest peaks were selected and all objects with
signal-to-noise values higher than 3--4, depending on number of
observational points, were visually examined. Due to limited number of
measurements power spectra were prone to display periodicities around 1-day
corresponding to observation sampling, so plenty of objects had to be
rejected. The remaining stars were classified into the main variability
types -- Cepheid variables, RR~Lyr variables, eclipsing binaries,
long-period variables -- and a group of miscellaneous variables.

\Section{Classical Cepheids}
\vspace*{-2pt}
In the main OGLE-III catalog (Soszyñski \etal 2008) 26 bright
fundamental-mode Cepheids in the LMC were saturated at least in one filter
while five objects were saturated in both filters. It prevented Soszyñski
\etal (2008) from including them in the complete period--luminosity
analysis. Only 21 objects from that sample in the {\it V}-band relation
were used and the relation shows significant scatter. The Shallow Survey
allowed us to obtain good quality {\it VI} photometry for these
Cepheids. Also two completely new Cepheid variables were found with
BRIGHT-CEP-1 object apparently belonging to the NGC\,1698 star
cluster. They were omitted in the main catalog due to the location
close to the saturated objects masked out on the OGLE-III reference image.
Fig.~1
%\vspace*{-15pt}
\begin{figure}[htb]
\centerline{\includegraphics[width=12.5cm]{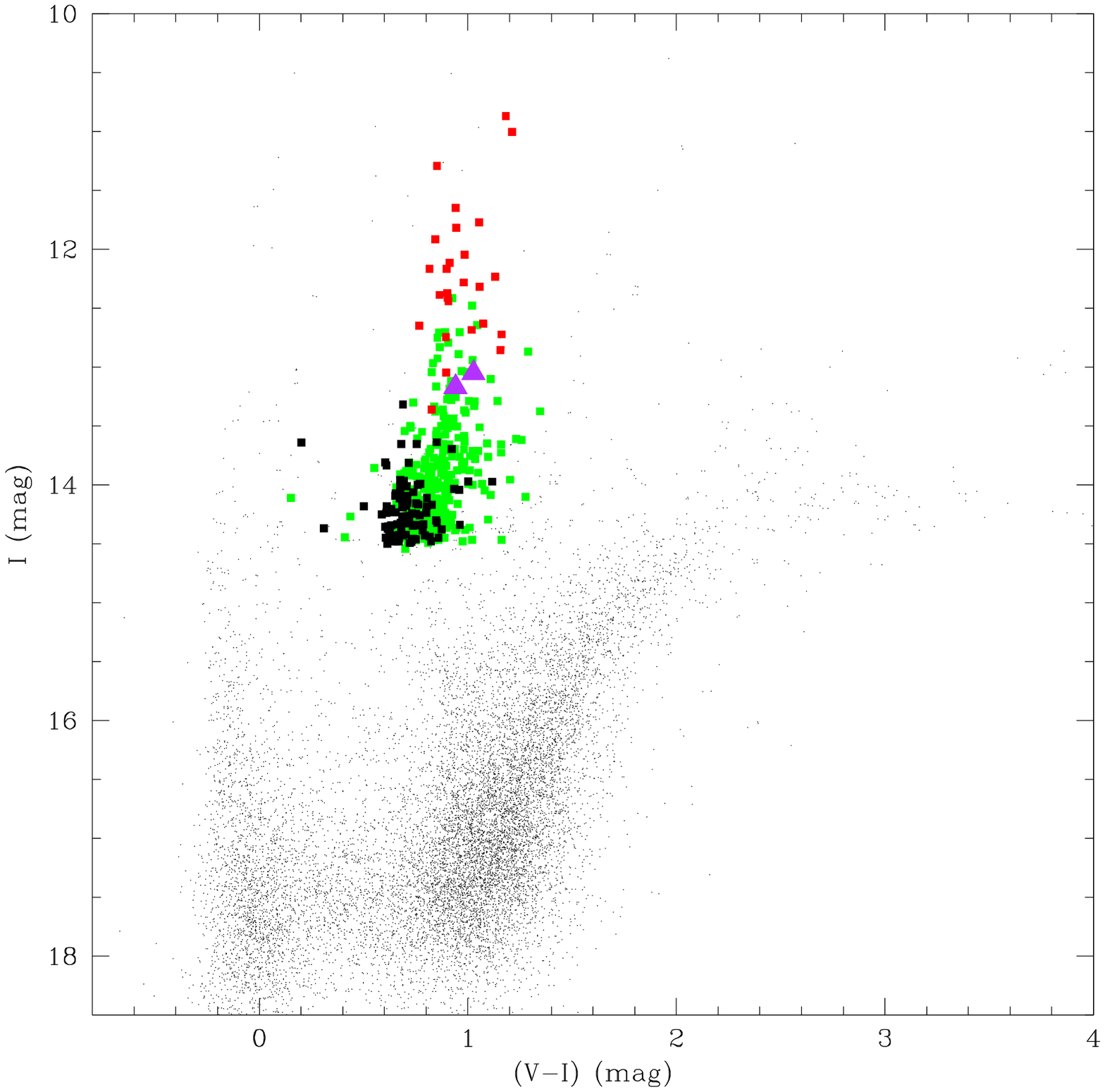}}
\FigCap{Color--magnitude diagram for Cepheids. Red points correspond to 
stars for which new photometry was obtained and violet triangles correspond to
newly found objects. Fundamental-mode pulsators from the OGLE-III catalog
are marked with green points and secondary-mode pulsators (mostly first
overtone) with black points. All measurements are from the Shallow
Survey.}
\end{figure}
presents the color--magnitude diagram for all Cepheids classified in the
survey with emphasizing the objects for which photometric data were not
available or were incomplete. We also show the {\it VI}-band light curves
in Figs.~2--5. The light curve of OGLE-LMC-CEP-0619 variable appears to
have additional structure. We subtracted dominant period in both bands but
the results are inconclusive. The comparison with the ASAS\footnote{\it
http://www.astrouw.edu.pl/asas/} (Pojmañski 2002) data suggests that its amplitude is
modulated.
\begin{figure}[htb]
\centerline{\includegraphics[width=13cm,clip,trim=0 30 0 30]{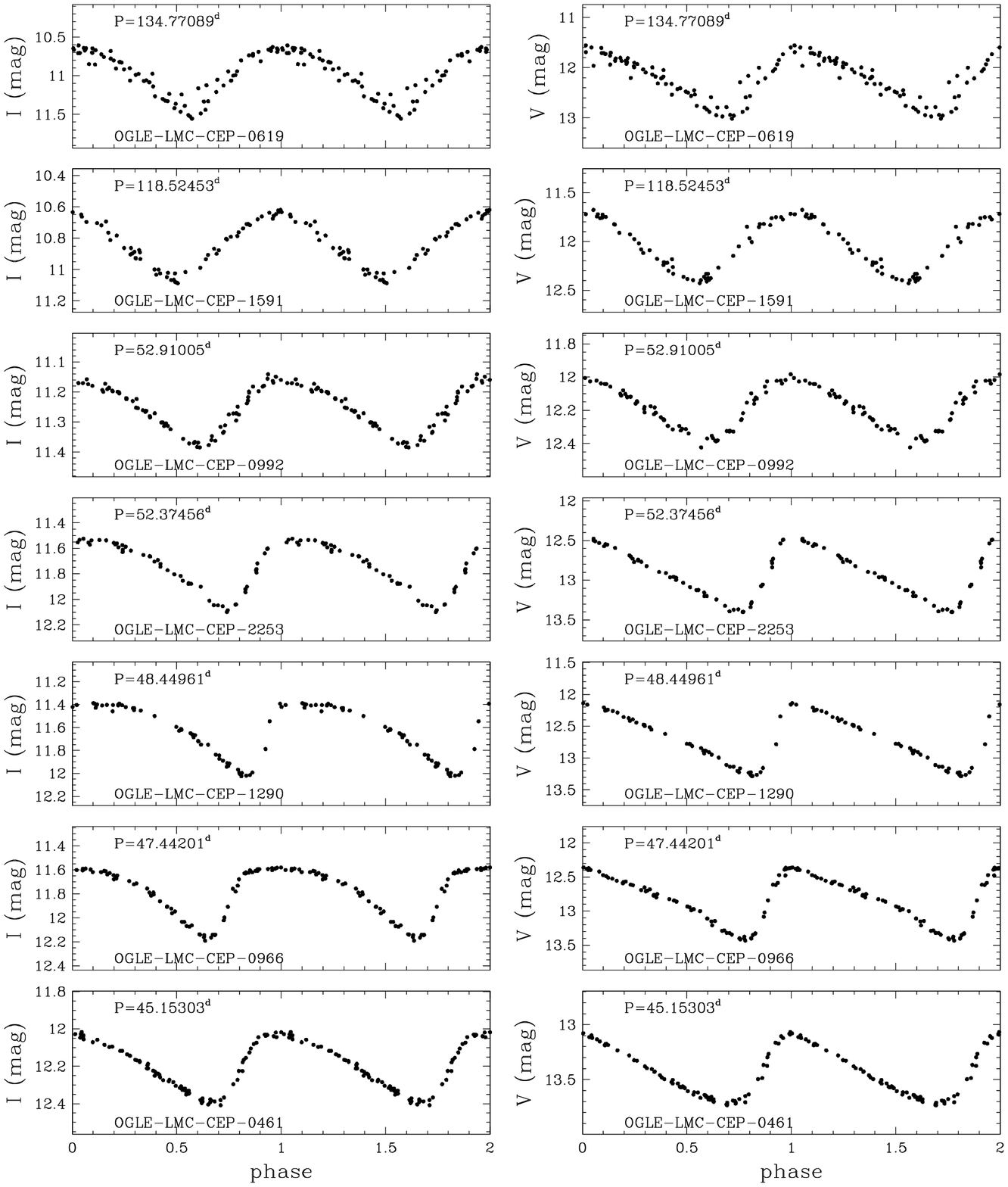}}
\vspace*{-5mm}
\FigCap{{\it VI}-band light curves of Cepheids for which new photometry 
was obtained, sorted according to decreasing period (periods range 135 to 45 days).}
\end{figure}
\begin{figure}[htb]
\vglue-6mm
\centerline{\includegraphics[width=13cm,clip,trim=0 30 0 0]{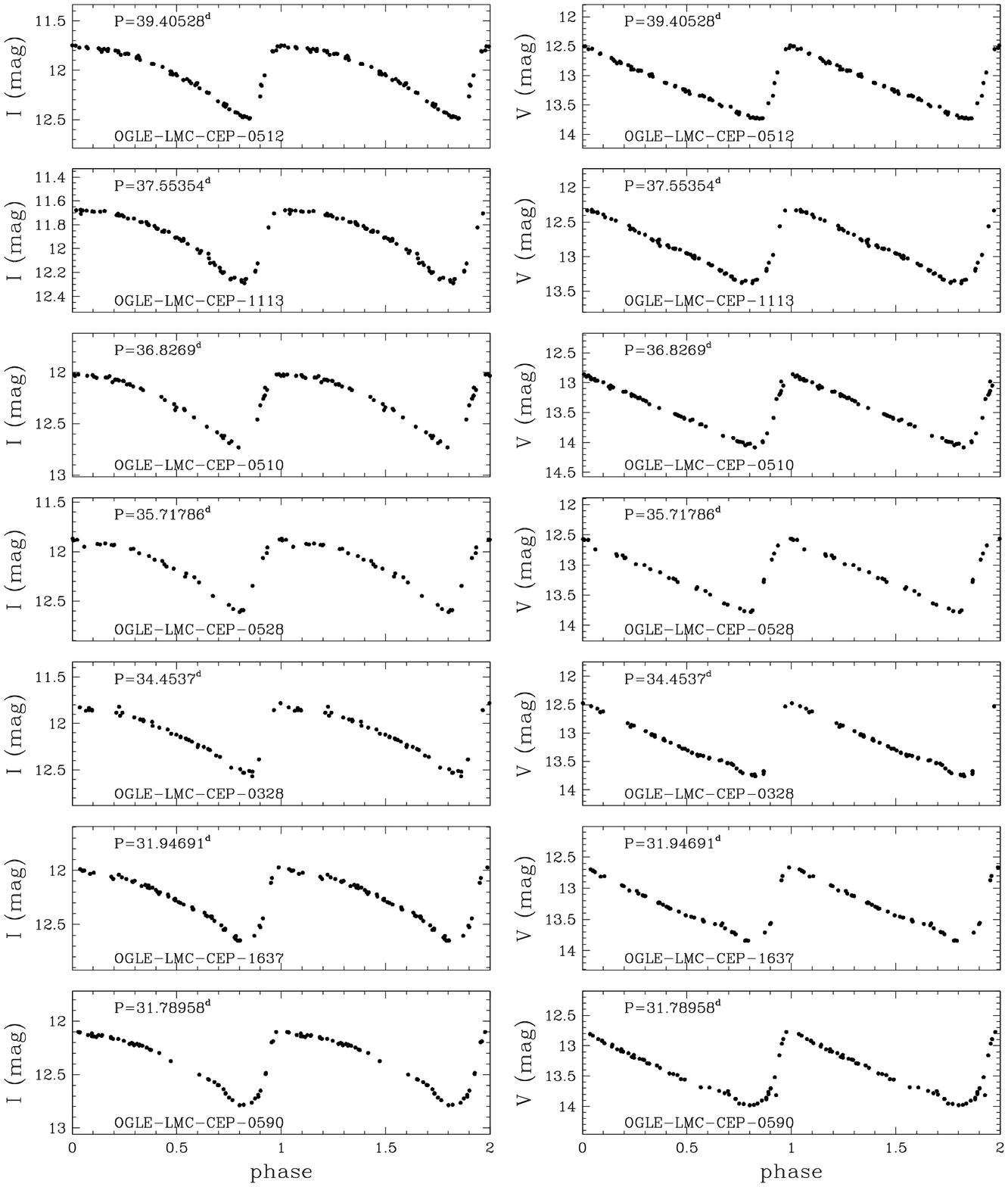}}
\vspace*{-6mm}
\FigCap{{\it VI}-band light curves of Cepheids for which new photometry 
was obtained, sorted according to decreasing period (periods range 40 to 31 days).}
\end{figure}

In the observed fields 11 of total 416 objects already present in the
OGLE-III catalogs in the range of analyzed magnitudes were overlooked by
the Shallow Survey search procedure mainly due to localization close to the
detector edge or small amplitudes of variability combined with limited number
of observational points. For each object Fourier series with eight
harmonics was fitted to the folded {\it VI} light curves and mean
magnitudes were derived by integration of light curves in intensity
units. Fig.~6 presents Fourier coefficients $R_{21}$, $R_{31}$, $\phi_{21}$
and $\phi_{31}$ as a function of the period, where amplitude ratios
$R_{k1}=A_k/A_1$ and phase differences $\phi_{k1}=\phi_{k}-k\phi_{1}$.
\begin{figure}[htb]
\vglue-7mm
\centerline{\includegraphics[width=13cm,clip,trim=0 30 0 30]{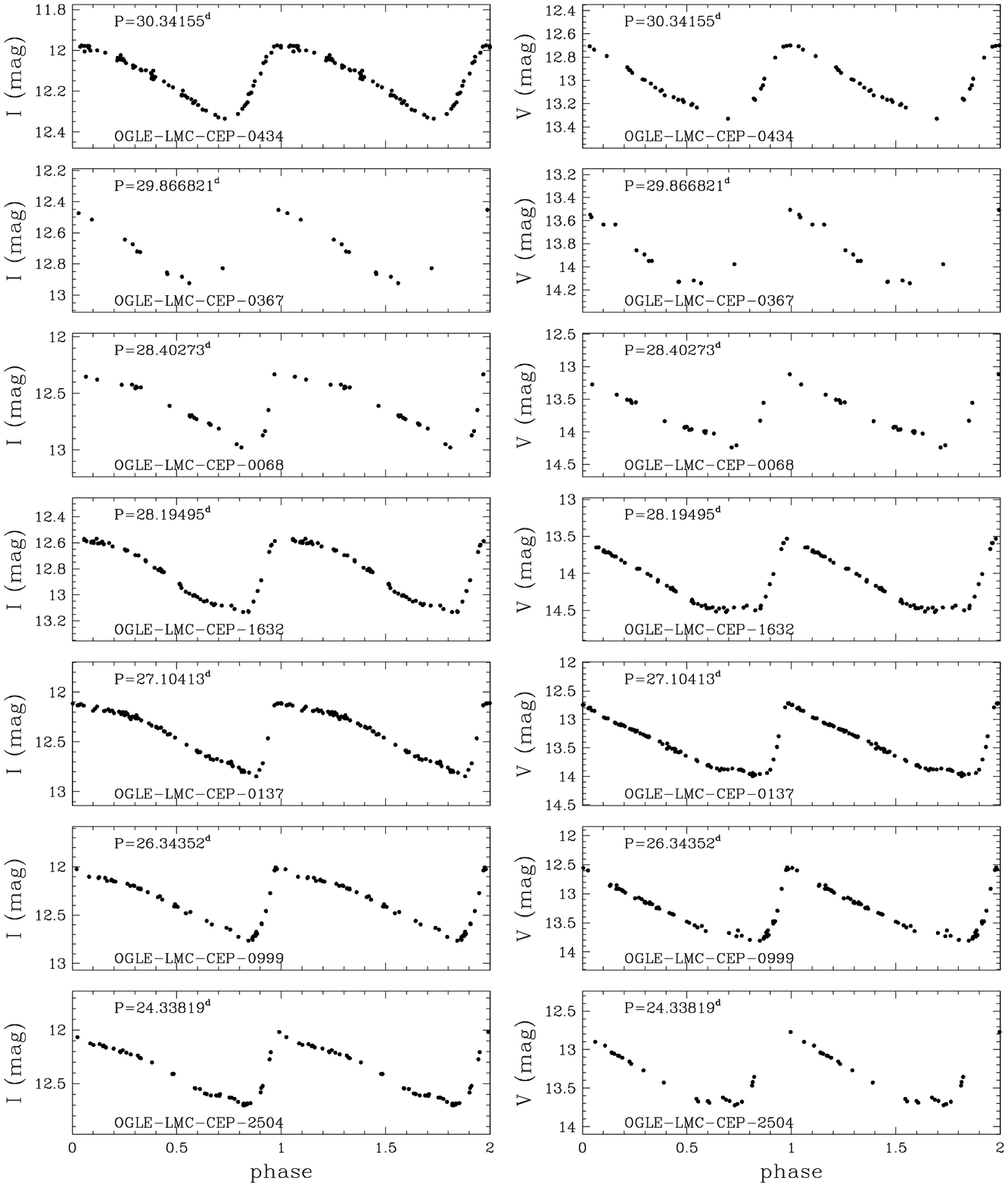}}
\vspace*{-5mm}
\FigCap{{\it VI}-band light curves of Cepheids for which new photometry 
was obtained, sorted according to decreasing period (periods range 31 to 24 days).}
\end{figure}

\vspace*{-1mm}
Table~1 contains basic parameters for each Cepheid. In the subsequent
columns the following data are presented:
(1)~variable star identification label (new objects have prefix ``BRIGHT''), 
(2,3)~equatorial coordinates J2000.0, 
(4)~subfield designation, 
(5)~period,
(6)~{\it I}-band mean magnitude,
(7)~{\it V}-band mean magnitude,
(8)~({\it V-I}) color index, 
(9-14)~Fourier parameters for the {\it I}-band curve, 
(15-20)~Fourier parameters for the {\it V}-band curve,
(21)~number of observations in the {\it I}-band, 
(22)~number of observations in the {\it V}-band.

\begin{figure}[htb]
\vglue-5mm
\centerline{\includegraphics[width=13.5cm,clip,trim=0 30 0 0]{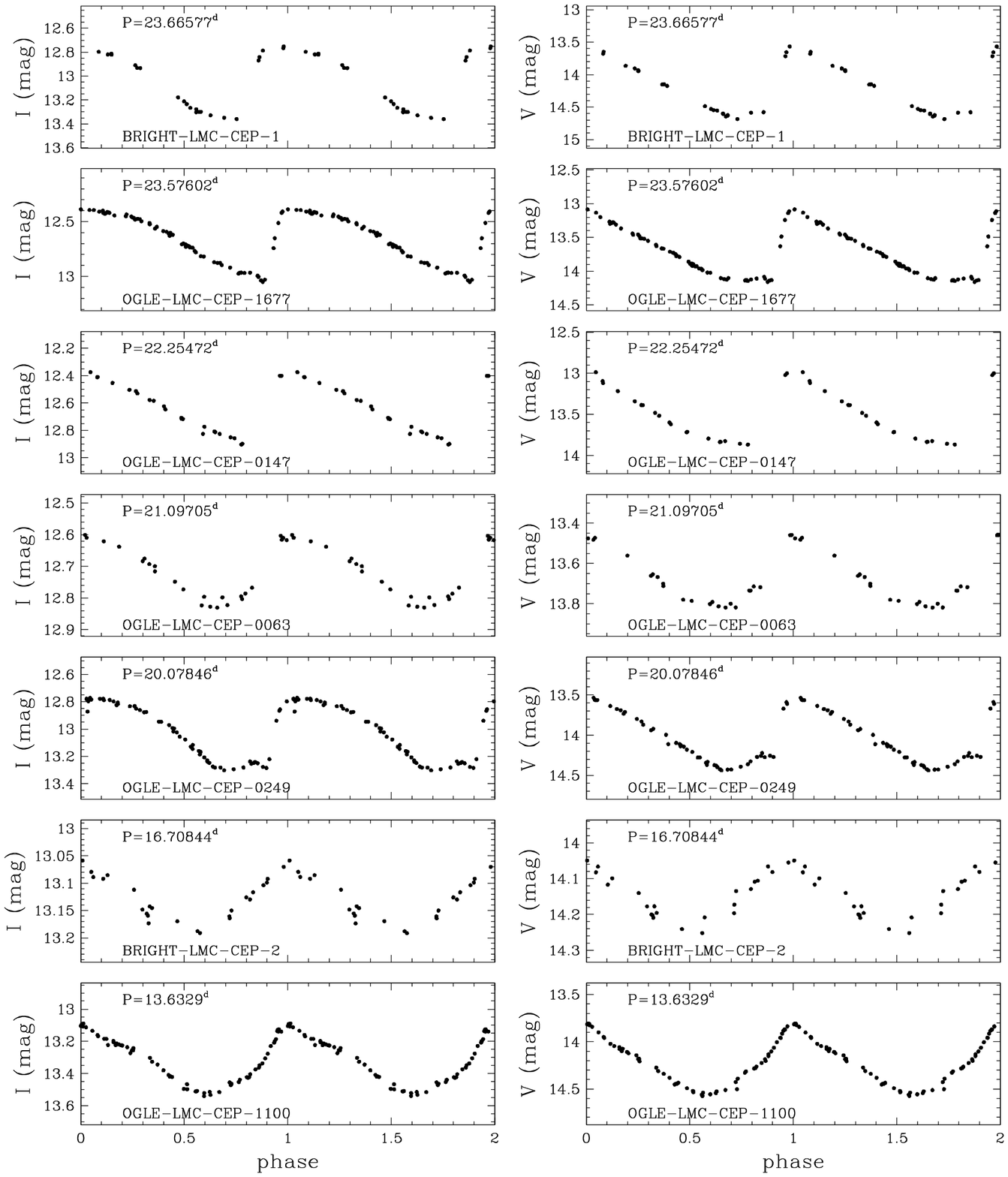}}
\vspace*{-5mm}
\FigCap{{\it VI}-band light curves of Cepheids for which new photometry 
was obtained, sorted according to decreasing period (periods range 24 to 13 days).}
\end{figure}
\begin{figure}[p]
%\centerline{\includegraphics[width=0.8\textwidth,clip,trim=0 15 0 30]{fig6a.ps}}
%\centerline{\includegraphics[width=0.8\textwidth]{fig6b.ps}}
\centerline{\includegraphics[width=9.7cm, bb=20 165 575 700]{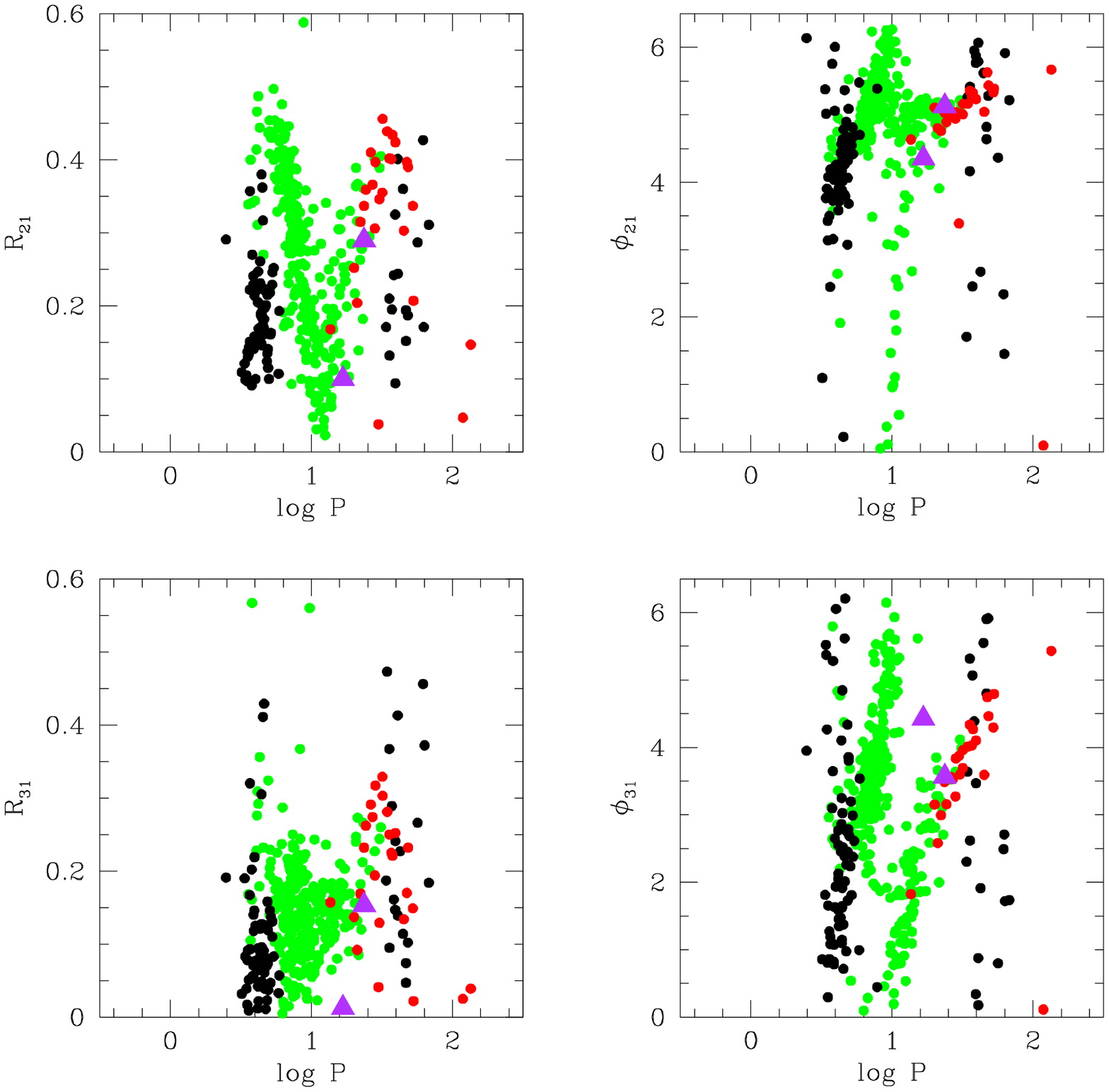}}
\centerline{\includegraphics[width=9.7cm, bb=20 165 575 700]{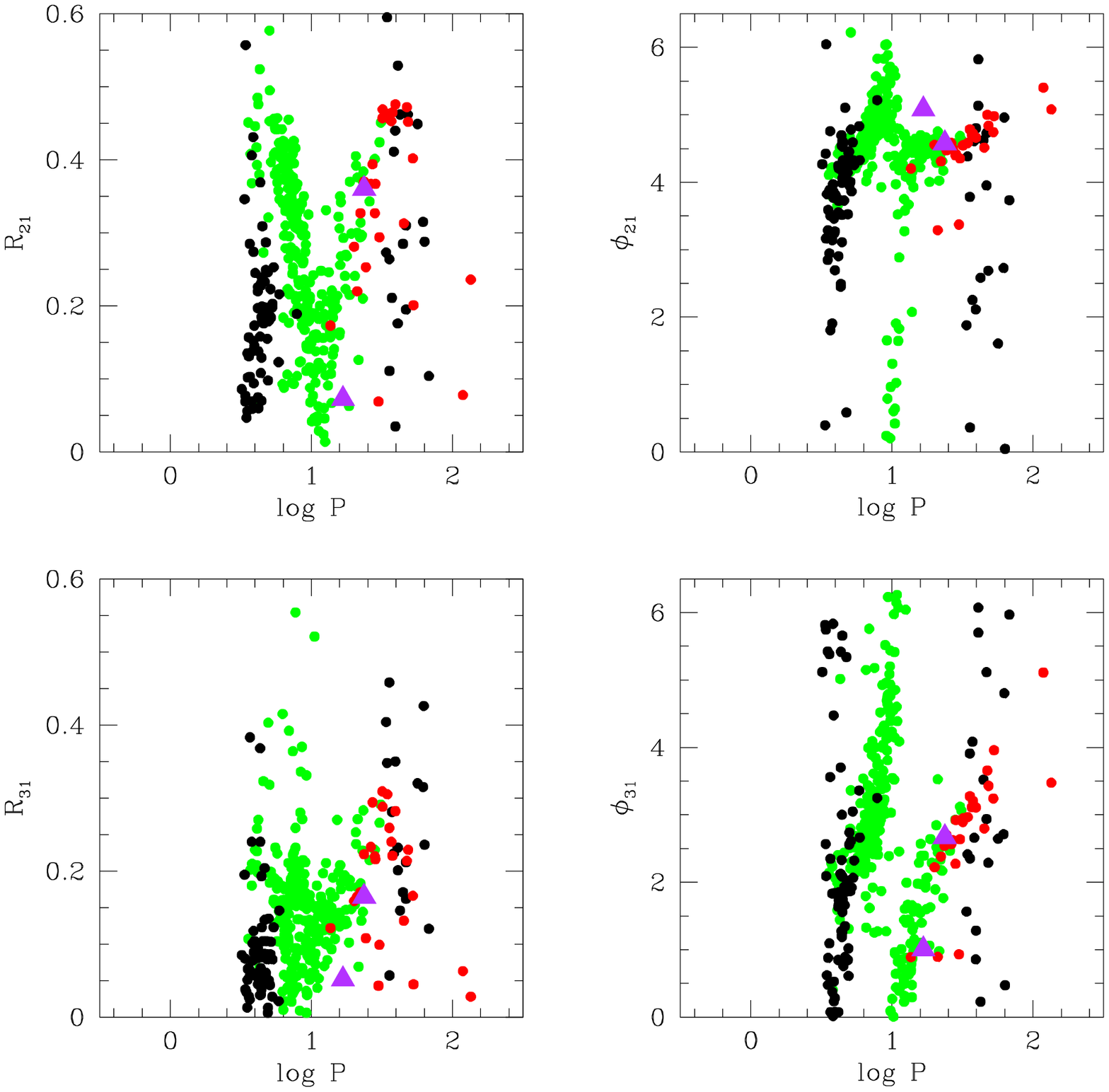}}
%\vspace*{-6pt}
\FigCap{Fourier parameters for Cepheids. Red points represent objects 
with new photometry and violet triangles are new Cepheids. Fundamental-mode
pulsators from the OGLE-III survey are marked with green color. Black dots
are Cepheids pulsating in non-fundamental modes. {\it Four upper panels}
are for {\it I}-band and the {\it lower ones} are for {\it V}-band. All
measurements are from the Shallow Survey.}
\end{figure}
For fundamental-mode pulsators with periods longer than 10 days $\log
P{-}I$ and $\log P{-}W_I$ relations were fitted using the least squares method
with $3\sigma$ clipping:
\begin{eqnarray}
W_I=-3.294(\pm0.050)\log P+15.847(\pm0.063) &\quad \sigma=0.118\\ \nonumber
  I=-2.984(\pm0.085)\log P+16.924(\pm0.108) &\quad \sigma=0.200 
\end{eqnarray}

\begin{landscape}
\renewcommand{\TableFont}{\tiny}
\renewcommand{\arraystretch}{1.25}
\MakeTableSep{
r@{\hspace{6pt}}
c@{\hspace{6pt}}
c@{\hspace{6pt}}
c@{\hspace{6pt}}
c@{\hspace{6pt}}
c@{\hspace{6pt}}
c@{\hspace{6pt}}
c@{\hspace{6pt}}
|c@{\hspace{6pt}}
c@{\hspace{6pt}}
c@{\hspace{6pt}}
c@{\hspace{6pt}}
c@{\hspace{6pt}}
c|@{\hspace{6pt}}
c@{\hspace{6pt}}
c@{\hspace{6pt}}
c@{\hspace{6pt}}
c@{\hspace{6pt}}
c@{\hspace{6pt}}
c|@{\hspace{6pt}}
c@{\hspace{6pt}}
c}
{22cm}{Basic parameters of Cepheids derived with OGLE Shallow Survey photometry}
{
\hline
\uprule
&&&&&&&&\multicolumn{5}{c}{\hspace{0.5cm}{\it I}-band light
curve}&&\multicolumn{5}{c}{\hspace{0.5cm}{\it V}-band light curve}&&&\\
ID\hspace{0.6cm} & RA & DEC & Subfield & Period & $I$ & $V$  & $V-I$ &
$T_{\rm max}$ & $A$ & $R_{21}$ & $\phi_{21}$ & $R_{31}$ & $\phi_{31}$ &
$T_{\rm max}$ & $A$ & $R_{21}$ & $\phi_{21}$ & $R_{31}$ & $\phi_{31}$ & $N_I$ & $N_V$ \\
\dorule
&(J2000.0) & (J2000.0)&  & [days] & [mag] & [mag] & [mag]
& \raisebox{0.05cm}{$_{\rm [HJD-2450000]}$} & & & & &
& \raisebox{0.05cm}{$_{\rm [HJD-2450000]}$} & & & & & & &\\
\hline
\uprule
BRIGHT-LMC-CEP-1 & 4\uph49\upm04.5\ups & -69\arcd06\arcm43\arcs & LMC142.3 &    23.66577 & 13.050 & 14.077 &  1.027&   4345.69600 & 0.60 & 0.289 & 5.113 & 0.158 & 3.550 &   4343.25225 & 1.08 & 0.360 & 4.582 & 0.165 & 2.671& 24 & 25\\
BRIGHT-LMC-CEP-2 & 4\uph46\upm08.4\ups & -69\arcd02\arcm47\arcs & LMC142.5 &    16.70844 & 13.169 & 14.110 &  0.941&   4359.42193 & 0.13 & 0.100 & 4.358 & 0.013 & 4.430 &   4359.49039 & 0.20 & 0.073 & 5.081 & 0.052 & 1.005& 25 & 24\\
OGLE-LMC-CEP-0063 & 4\uph45\upm18.7\ups & -69\arcd16\arcm38\arcs & LMC142.7 &    21.09705 & 12.745 & 13.624 &  0.879&   4348.21410 & 0.21 & 0.204 & 4.804 & 0.092 & 2.581 &   4326.83017 & 0.35 & 0.195 & 3.968 & 0.099 & 2.025& 25 & 23\\
OGLE-LMC-CEP-0068 & 4\uph45\upm37.2\ups & -70\arcd15\arcm04\arcs & LMC144.4 &    28.40273 & 12.633 & 13.706 &  1.073&   4359.54071 & 0.65 & 0.397 & 5.038 & 0.317 & 3.837 &   4333.19359 & 1.08 & 0.366 & 4.421 & 0.215 & 2.918& 25 & 23\\
OGLE-LMC-CEP-0137 & 4\uph49\upm53.8\ups & -69\arcd45\arcm16\arcs & LMC136.6 &    27.10413 & 12.441 & 13.348 &  0.907&   3231.96866 & 0.78 & 0.366 & 4.980 & 0.274 & 3.534 &   3231.88067 & 1.33 & 0.394 & 4.496 & 0.294 & 2.623& 77 & 76\\
OGLE-LMC-CEP-0147 & 4\uph50\upm03.2\ups & -68\arcd15\arcm34\arcs & LMC140.1 &    22.25472 & 12.649 & 13.416 &  0.767&   4332.13835 & 0.52 & 0.315 & 4.763 & 0.169 & 2.995 &   4354.40310 & 0.92 & 0.327 & 4.311 & 0.171 & 2.381& 25 & 21\\
OGLE-LMC-CEP-0249 & 4\uph52\upm47.7\ups & -68\arcd20\arcm56\arcs & LMC133.8 &    20.07846 & 13.048 & 13.946 &  0.898&   3223.26276 & 0.55 & 0.252 & 5.108 & 0.137 & 3.155 &   3223.11134 & 0.94 & 0.281 & 4.553 & 0.153 & 2.226& 57 & 47\\
OGLE-LMC-CEP-0328 & 4\uph54\upm23.9\ups & -70\arcd54\arcm05\arcs & LMC130.6 &    34.45370 & 12.114 & 13.027 &  0.913&   3275.00059 & 0.75 & 0.439 & 5.163 & 0.281 & 4.012 &   3309.25352 & 1.30 & 0.463 & 4.582 & 0.305 & 2.966& 44 & 50\\
OGLE-LMC-CEP-0367 & 4\uph55\upm06.3\ups & -67\arcd28\arcm33\arcs & LMC132.2 &    29.86682 & 12.734 & 13.890 &  1.156&   4338.43101 & 0.46 & 0.111 & 4.368 & 0.055 & 1.220 &   4338.21500 & 0.62 & 0.095 & 3.681 & 0.077 & 0.641& 12 & 14\\
OGLE-LMC-CEP-0434 & 4\uph56\upm27.5\ups & -69\arcd22\arcm46\arcs & LMC127.7 &    30.34155 & 12.164 & 12.980 &  0.816&   3235.06302 & 0.36 & 0.346 & 5.016 & 0.129 & 3.598 &   3295.83287 & 0.63 & 0.294 & 4.354 & 0.099 & 2.639& 69 & 32\\
OGLE-LMC-CEP-0461 & 4\uph57\upm01.8\ups & -67\arcd59\arcm43\arcs & LMC133.3 &    45.15303 & 12.232 & 13.363 &  1.131&   3202.10571 & 0.36 & 0.303 & 5.045 & 0.134 & 3.593 &   3198.96320 & 0.64 & 0.313 & 4.513 & 0.132 & 2.798& 86 & 74\\
OGLE-LMC-CEP-0510 & 4\uph58\upm05.6\ups & -69\arcd27\arcm15\arcs & LMC127.8 &    36.82690 & 12.318 & 13.374 &  1.056&   3225.80352 & 0.72 & 0.401 & 5.260 & 0.225 & 4.028 &   3224.73503 & 1.20 & 0.453 & 4.705 & 0.240 & 3.114& 50 & 63\\
OGLE-LMC-CEP-0512 & 4\uph58\upm10.8\ups & -69\arcd56\arcm59\arcs & LMC128.7 &    39.40528 & 12.046 & 13.031 &  0.985&   3224.19225 & 0.73 & 0.424 & 5.233 & 0.252 & 4.101 &   3223.68992 & 1.24 & 0.476 & 4.661 & 0.282 & 3.109& 73 & 69\\
OGLE-LMC-CEP-0528 & 4\uph58\upm32.8\ups & -70\arcd20\arcm45\arcs & LMC129.3 &    35.71786 & 12.166 & 13.064 &  0.898&   3298.51189 & 0.73 & 0.402 & 5.356 & 0.250 & 4.337 &   3369.78102 & 1.22 & 0.456 & 4.785 & 0.259 & 3.277& 35 & 34\\
OGLE-LMC-CEP-0590 & 4\uph59\upm41.2\ups & -69\arcd27\arcm21\arcs & LMC127.1 &    31.78958 & 12.411 & 13.314 &  0.903&   3231.96580 & 0.73 & 0.355 & 5.006 & 0.329 & 3.694 &   3327.33092 & 1.30 & 0.456 & 4.540 & 0.309 & 2.881& 51 & 54\\
OGLE-LMC-CEP-0619 & 5\uph00\upm07.6\ups & -68\arcd26\arcm60\arcs & LMC126.4 &   134.77089 & 11.005 & 12.218 &  1.213&   3235.04996 & 0.80 & 0.147 & 5.670 & 0.039 & 5.431 &   3214.64903 & 1.29 & 0.236 & 5.080 & 0.028 & 3.476& 77 & 71\\
OGLE-LMC-CEP-0966 & 5\uph06\upm48.0\ups & -70\arcd02\arcm13\arcs & LMC120.1 &    47.44201 & 11.817 & 12.761 &  0.944&   3228.24951 & 0.58 & 0.397 & 5.629 & 0.170 & 4.746 &   3221.44380 & 1.05 & 0.472 & 5.000 & 0.214 & 3.657& 69 & 70\\
OGLE-LMC-CEP-0992 & 5\uph07\upm16.0\ups & -68\arcd53\arcm01\arcs & LMC118.1 &    52.91005 & 11.294 & 12.146 &  0.852&   3230.20460 & 0.22 & 0.207 & 5.390 & 0.022 & 4.792 &   3228.20959 & 0.40 & 0.201 & 4.980 & 0.045 & 3.959& 73 & 65\\
OGLE-LMC-CEP-0999 & 5\uph07\upm20.1\ups & -70\arcd27\arcm15\arcs & LMC121.2 &    26.34352 & 12.371 & 13.272 &  0.901&   3218.57167 & 0.77 & 0.410 & 4.973 & 0.291 & 3.561 &   3218.40084 & 1.33 & 0.367 & 4.582 & 0.233 & 2.564& 49 & 57\\
OGLE-LMC-CEP-1100 & 5\uph09\upm04.4\ups & -70\arcd21\arcm54\arcs & LMC113.6 &    13.63290 & 13.359 & 14.185 &  0.826&   3224.43455 & 0.44 & 0.167 & 4.636 & 0.158 & 1.822 &   3238.02595 & 0.76 & 0.173 & 4.203 & 0.122 & 0.890& 67 & 63\\
OGLE-LMC-CEP-1113 & 5\uph09\upm20.1\ups & -70\arcd27\arcm27\arcs & LMC113.7 &    37.55354 & 11.917 & 12.760 &  0.843&   3203.78032 & 0.61 & 0.434 & 5.326 & 0.221 & 4.270 &   3203.60447 & 1.09 & 0.465 & 4.724 & 0.221 & 3.204& 66 & 64\\
OGLE-LMC-CEP-1290 & 5\uph13\upm53.7\ups & -67\arcd03\arcm49\arcs & LMC107.1 &    48.44961 & 11.648 & 12.590 &  0.942&   3221.34724 & 0.62 & 0.390 & 5.435 & 0.232 & 4.464 &   3221.23847 & 1.12 & 0.452 & 4.838 & 0.229 & 3.430& 52 & 47\\
OGLE-LMC-CEP-1591 & 5\uph19\upm30.5\ups & -68\arcd41\arcm10\arcs & LMC101.2 &   118.52453 & 10.870 & 12.052 &  1.182&   3234.66069 & 0.43 & 0.047 & 0.095 & 0.025 & 0.113 &   3221.73250 & 0.70 & 0.078 & 5.403 & 0.063 & 5.110& 57 & 54\\
OGLE-LMC-CEP-1632 & 5\uph20\upm17.5\ups & -67\arcd56\arcm53\arcs & LMC102.3 &    28.19495 & 12.856 & 14.012 &  1.156&   3230.04289 & 0.59 & 0.306 & 4.945 & 0.194 & 3.273 &   3229.76823 & 1.00 & 0.327 & 4.400 & 0.220 & 2.274& 57 & 57\\
OGLE-LMC-CEP-1637 & 5\uph20\upm23.0\ups & -69\arcd02\arcm18\arcs & LMC100.4 &    31.94691 & 12.282 & 13.261 &  0.979&   3212.29806 & 0.71 & 0.456 & 5.162 & 0.303 & 3.962 &   3212.24026 & 1.21 & 0.469 & 4.548 & 0.288 & 2.948& 63 & 48\\
OGLE-LMC-CEP-1677 & 5\uph21\upm12.5\ups & -69\arcd03\arcm08\arcs & LMC100.4 &    23.57602 & 12.685 & 13.704 &  1.019&   3220.20153 & 0.67 & 0.337 & 5.035 & 0.232 & 3.489 &   3220.07714 & 1.09 & 0.370 & 4.563 & 0.223 & 2.541& 65 & 62\\
OGLE-LMC-CEP-2253 & 5\uph31\upm21.8\ups & -70\arcd57\arcm25\arcs & LMC171.6 &    52.37456 & 11.772 & 12.826 &  1.054&   3232.39835 & 0.55 & 0.337 & 5.333 & 0.149 & 4.297 &   3231.01078 & 0.99 & 0.402 & 4.743 & 0.166 & 3.243& 50 & 49\\
\dorule
OGLE-LMC-CEP-2504 & 5\uph35\upm36.1\ups & -68\arcd32\arcm04\arcs & LMC174.6 &    24.33819 & 12.388 & 13.240 &  0.852&   3237.71028 & 0.68 & 0.359 & 4.894 & 0.262 & 3.157 &   3239.88429 & 0.91 & 0.280 & 4.342 & 0.111 & 2.241& 47 & 26\\
\hline
}
\end{landscape}

\begin{figure}[htb]
\vglue-7mm
\centerline{\includegraphics[width=13cm]{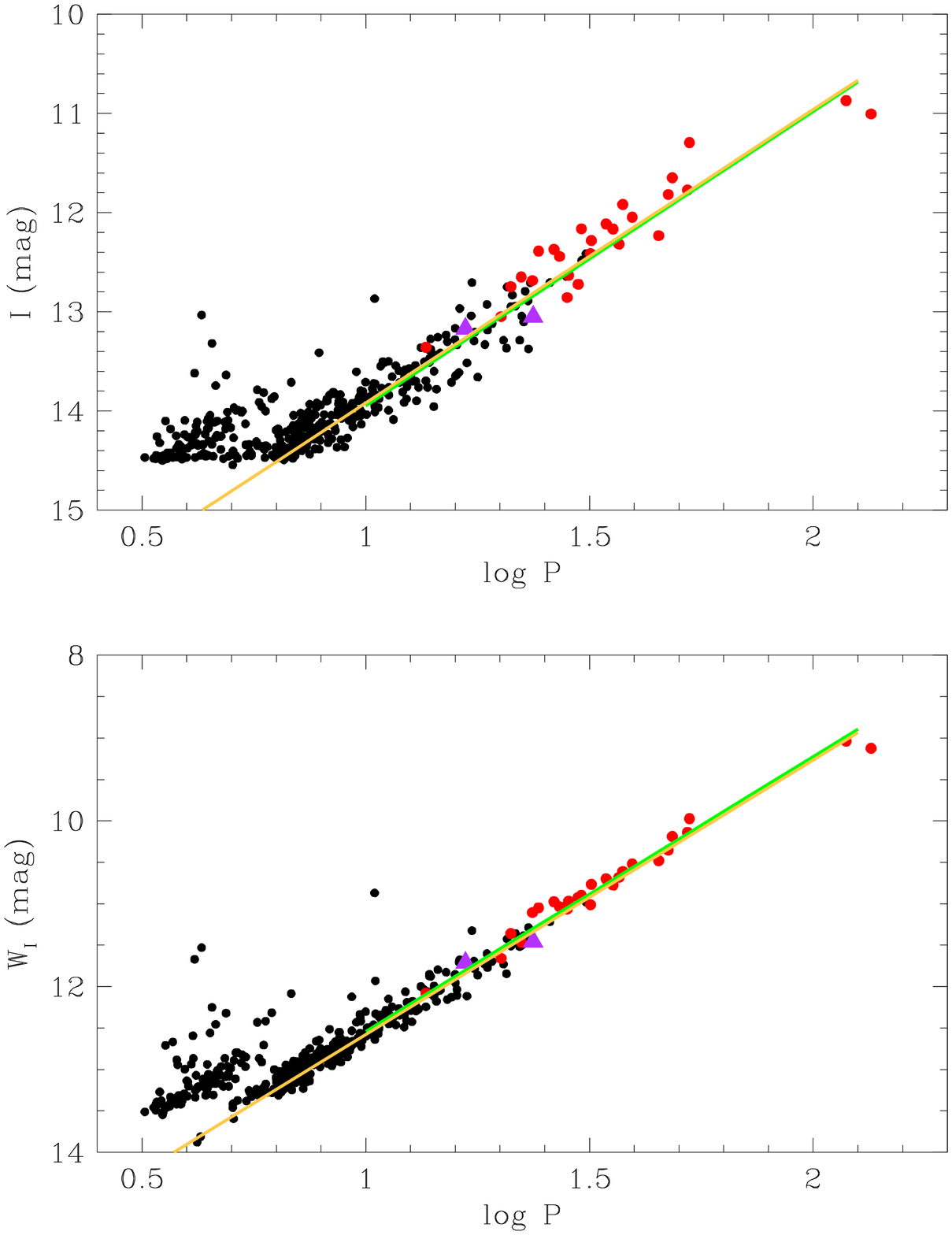}}
\vspace*{-5mm}
\FigCap{Period--luminosity diagrams for classical Cepheids. Red points 
correspond to stars for which new photometry was obtained and violet triangles
correspond to newly found objects. Cepheids from the main OGLE-III catalog
are marked with black dots. Green line shows the fit for Cepheids with
periods above 10 days and yellow line represents relation derived in
Soszyñski \etal (2008). All measurements are from the Shallow Survey.}
\end{figure}
We used Wesenheit index (Madore 1982) defined as:
$$W_I=I-1.55\cdot(V-I).\eqno(2)$$

Relations presented in Soszyñski \etal (2008) for Cepheids of periods below
$\log P=1.5$ were as follows:
\setcounter{equation}{2}
\begin{eqnarray}
W_I=-3.314(\pm0.009)\log P+15.893(\pm0.006) & \quad\sigma=0.080 \\ \nonumber
  I=-2.959(\pm0.016)\log P+16.879(\pm0.010) & \quad\sigma=0.150 
\end{eqnarray}

The results are graphically presented in Fig.~7. Red points correspond to
the Cepheids for which new photometry was obtained and newly detected
Cepheids are marked with violet triangles. The green line is the relation
for Cepheids with periods longer than 10 days and the yellow one is the
relation derived in Soszyñski \etal (2008) paper. One can note that there
are no statistically significant deviations in relations for longer period
domain. Obviously, we have limited sample of long-period Cepheids in the
LMC and, moreover, we did not detect any variables of this type with
periods ranging from about 55~days to 115~days. 

It is worth noting that mean magnitude scatter for brighter Cepheids is
smaller than in the diagrams based on the main OGLE-III photometry
(Soszyñski \etal 2008). This effect, specifically visible for $\log P>1.2$,
can be caused by non-linear CCD response close to the saturation limit
and/or by imperfections in DIA subtraction procedures for luminous
stars. We also checked that significant color index shifts for some objects
are caused by stronger blending in the Shallow Survey images in comparison
to the regular OGLE-III data.
\vspace*{2mm}
\Section{RR~Lyr Variables}
In the magnitude range of the Shallow Survey searched for variables, the
main OGLE-III catalog (Soszyñski \etal 2009a) contains ten RR~Lyr stars
located in the observed fields. We identified one new RR~Lyr variable at
equatorial coordinates: ${\rm RA=5\uph03\upm01\zdot\ups3}$, ${\rm
DEC}=-69\arcd09\arcm01\arcs$.  Considering its brightness it must be a
foreground Galactic object. Despite of the fact that the star is saturated
in the OGLE-III database we were able to calculate its proper motion based
on the extracted {\sc DoPhot} positions data derived in the main OGLE-III
survey :
\begin{eqnarray*}
\mu_{\delta}&=&~~8.17~\pm~0.32~{\rm mas/yr}\\
\mu_{\alpha}\cos\delta&=&10.06~\pm~0.32~{\rm mas/yr}\nonumber
\end{eqnarray*}

We applied the method presented by Poleski \etal (2012). The proper motion
was appropriately corrected so the resultant proper motion of all LMC stars
was equal 0 with an error of $0.25$ mas/yr. No parallax effect was visible.

Also we added to our catalog two RR~Lyr stars which were already known --
both are present in the ASAS catalog (Pojmañski 2002) and one also has
XX~Dor designation. The basic data for new RR~Lyr stars are presented in
the OGLE Internet Archive in table of the same format as for Cepheids. The
last extra column contains cross-identification designations. Fig.~8 shows
{\it I}- and {\it V}-band light curves for those stars.
\begin{figure}[htb]
\centerline{\includegraphics[width=12.5cm,clip, trim=5 310 0 30]{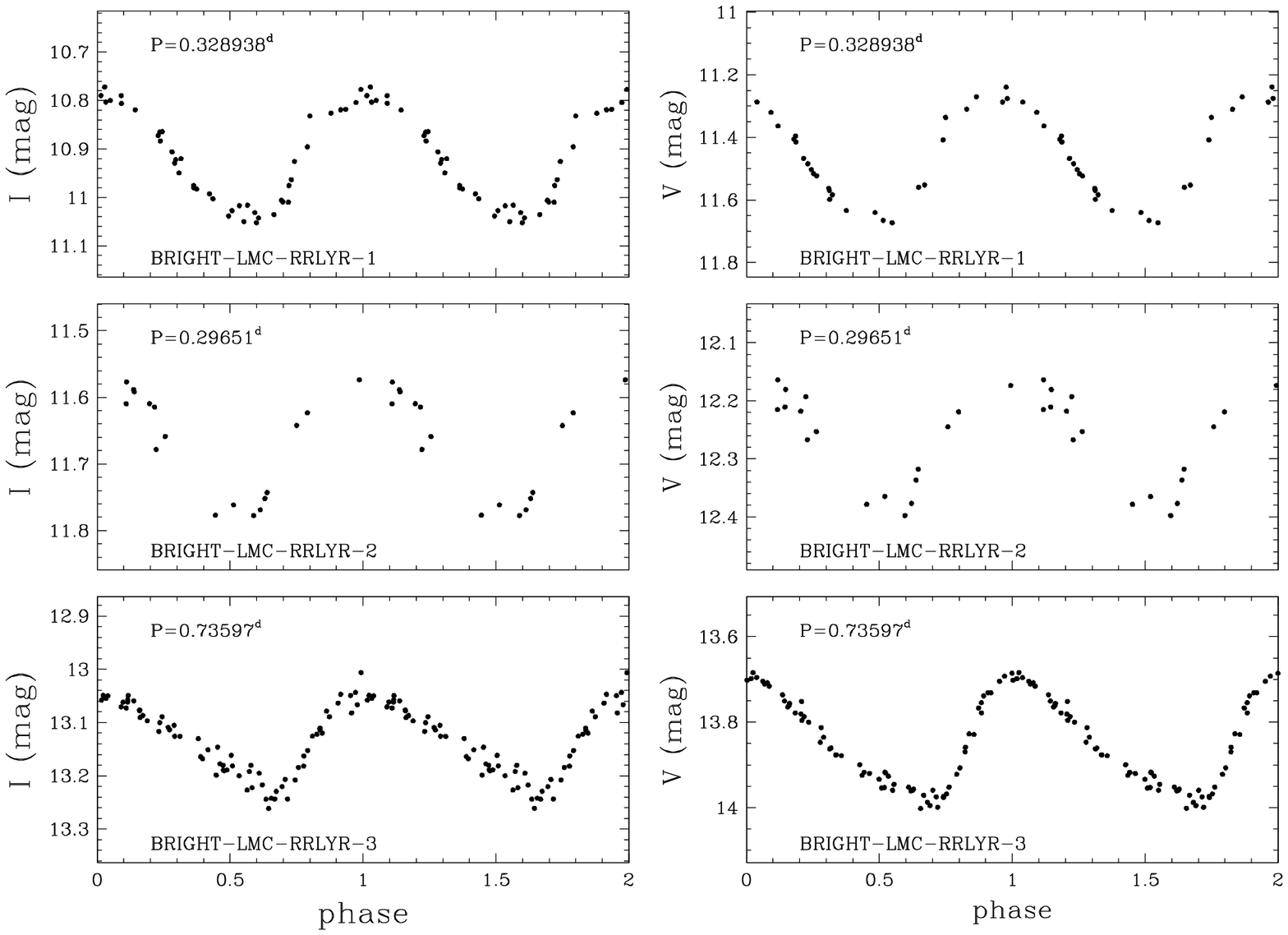}}
\vskip6pt
\FigCap{New Galactic RR~Lyr variables added to the OGLE-III catalog.}
\vglue3mm
\end{figure}
\Section{Eclipsing Variables}
We updated eclipsing variables catalog with 82 new objects. The main
OGLE-III catalog was focused on detached binaries (Graczyk \etal 2011) so
it is not surprising that we found many contact (EW) or semi-detached
systems (EB). We also discovered 32 ellipsoidal variables for which
eclipses are observed or not. Six objects were already present in the ASAS
catalog (Szczygie³ \etal 2010).

Table~2 lists new eclipsing binaries. Subsequent columns contain the
following data:
(1)~variable star identification label, 
(2,3)~equatorial coordinates J2000.0, 
(4)~subfield designation, 
(5)~period,
(6)~{\it I}-band mean magnitude,
(7)~{\it V}-band mean magnitude, 
(8)~({\it V-I}) color index, 
(9,10)~number of observations in the {\it I}- and {\it V}-band, 
(11,12)~dispersion of magnitudes for the {\it I}- and {\it V}-band,
(13)~subtype,
(14)~designations in other catalogs/surveys: 
ASAS (no prefix, Pojmañski 2002), 
2MASS (Cutri \etal 2003), 
HV (Payne-Gaposchkin 1971),
HD (Cannon 1925),
Spitzer -- SSTISAGE1C (Vijh \etal 2009)) and SSTISAGEMC (Meixner \etal 2006), 
MACHO (Ochsenbein, Bauer and Marcout 2000),
UCAC2 (Zacharias \etal 2004),
[BE74] (Bohannan and Epps 1974),
[FBM2009] (Fari{\~n}a \etal 2009),
[L72] (Massey 2000),
NGC 2070 MEL (Melnick 1985). 
The subtypes were assigned based on light curve shape. Fig.~9 presents
color--magnitude diagram with positions of detected eclipsing binaries. 

\begin{landscape}
\renewcommand{\TableFont}{\scriptsize}
\renewcommand{\arraystretch}{1.05}
\MakeTableSep{
c@{\hspace{6pt}}
c@{\hspace{6pt}}
c@{\hspace{6pt}}
c@{\hspace{6pt}}
c@{\hspace{6pt}}
c@{\hspace{6pt}}
c@{\hspace{6pt}}
c@{\hspace{6pt}}
c@{\hspace{5pt}}
c@{\hspace{5pt}}
c@{\hspace{5pt}}
c@{\hspace{2pt}}
c@{\hspace{0pt}}
c
}
{22cm}{The first 25 lines of eclipsing variable stars catalog}{
\hline
\noalign{\vskip3pt}
ID & RA & DEC & Subfield  & \multicolumn{1}{c}{Period} & {\it I} & {\it V}  & $V-I$ & $N_I$ & $N_V$ & $\sigma_I$ & $\sigma_V$ & Subtype&Other\\
&(J2000.0) &(J2000.0) & &\multicolumn{1}{c}{[days]} & [mag] & [mag] & [mag]  & & & [mag]&[mag] & & designations\\
\noalign{\vskip3pt}
\hline
\noalign{\vskip3pt}
BRIGHT-LMC-ECL-01 & 5\uph18\upm32\zdot\ups6 & $ -68\arcd13\arcm32\arcs $ & LMC102.8 &    0.285464 & 10.133 & 11.108 & $  0.975 $ & 45 & 55 & 0.188 & 0.203 & EW & 051832-6813.6\\
BRIGHT-LMC-ECL-02 & 4\uph59\upm54\zdot\ups6 & $ -70\arcd48\arcm45\arcs $ & LMC130.4 &    0.406400 & 13.109 & 13.724 & $  0.616 $ & 47 & 59 & 0.072 & 0.065 & EB & --\\
BRIGHT-LMC-ECL-03 & 5\uph28\upm32\zdot\ups5 & $ -68\arcd36\arcm14\arcs $ & LMC167.6 &    0.576860 & 10.473 & 11.336 & $  0.863 $ & 43 & 39 & 0.048 & 0.040 & EB/EA & 052833-6836.2\\
BRIGHT-LMC-ECL-04 & 4\uph59\upm05\zdot\ups8 & $ -69\arcd31\arcm50\arcs $ & LMC127.8 &    0.579960 & 13.302 & 13.746 & $  0.444 $ & 32 & 43 & 0.059 & 0.060 & EW & --\\
BRIGHT-LMC-ECL-05 & 6\uph05\upm22\zdot\ups0 & $ -69\arcd47\arcm00\arcs $ & LMC206.6 &    0.593020 & 11.690 & 12.244 & $  0.554 $ & 17 & 17 & 0.076 & 0.077 & EW & 060521-6947.1\\
BRIGHT-LMC-ECL-06 & 5\uph55\upm36\zdot\ups1 & $ -70\arcd13\arcm07\arcs $ & LMC192.4 &    0.789600 & 12.344 & 13.021 & $  0.678 $ & 50 & 40 & 0.049 & 0.057 & EB & --\\
BRIGHT-LMC-ECL-07 & 5\uph18\upm25\zdot\ups6 & $ -69\arcd12\arcm13\arcs $ & LMC100.6 &    1.062028 & 14.449 & 14.247 & $ -0.202 $ & 64 & 59 & 0.056 & 0.052 & EW & 2MASS J05182564-6912128\\
BRIGHT-LMC-ECL-08 & 5\uph35\upm02\zdot\ups1 & $ -68\arcd43\arcm45\arcs $ & LMC174.7 &    1.086410 & 12.048 & 12.152 & $  0.103 $ & 57 & 58 & 0.095 & 0.047 & EA & 053503-6843.7\\
BRIGHT-LMC-ECL-09 & 5\uph38\upm28\zdot\ups5 & $ -69\arcd11\arcm19\arcs $ & LMC175.6 &    1.124180 & 14.385 & 14.485 & $  0.100 $ & 36 & 28 & 0.094 & 0.105 & EW & --\\
BRIGHT-LMC-ECL-10 & 5\uph10\upm27\zdot\ups1 & $ -67\arcd54\arcm56\arcs $ & LMC109.6 &    1.127451 & 14.130 & 13.967 & $ -0.162 $ & 30 & 31 & 0.061 & 0.079 & EW & HV 5623\\
BRIGHT-LMC-ECL-11 & 4\uph55\upm11\zdot\ups4 & $ -69\arcd22\arcm31\arcs $ & LMC135.2 &    1.238320 & 14.331 & 14.139 & $ -0.192 $ & 67 & 64 & 0.037 & 0.038 & EW & --\\
BRIGHT-LMC-ECL-12 & 5\uph10\upm29\zdot\ups2 & $ -69\arcd19\arcm16\arcs $ & LMC111.7 &    1.251180 & 13.794 & 14.273 & $  0.479 $ & 30 &  8 & 0.446 & 0.909 & EB & --\\
BRIGHT-LMC-ECL-13 & 5\uph35\upm17\zdot\ups9 & $ -68\arcd52\arcm50\arcs $ & LMC174.8 &    1.255580 & 14.190 & 15.258 & $  1.068 $ & 47 & 19 & 0.056 & 0.149 & EB & MACHO 82.8894.18\\
BRIGHT-LMC-ECL-14 & 5\uph37\upm30\zdot\ups9 & $ -69\arcd11\arcm07\arcs $ & LMC175.6 &    1.327192 & 14.011 & 14.617 & $  0.607 $ & 59 & 54 & 0.052 & 0.053 & EW & 2MASS J05373092-6911070\\
BRIGHT-LMC-ECL-15 & 5\uph04\upm51\zdot\ups5 & $ -70\arcd42\arcm36\arcs $ & LMC121.8 &    1.383040 & 14.293 & 14.088 & $ -0.205 $ & 69 & 66 & 0.021 & 0.020 & Ell & --\\
BRIGHT-LMC-ECL-16 & 5\uph27\upm54\zdot\ups1 & $ -68\arcd59\arcm45\arcs $ & LMC161.4 &    1.396120 & 14.384 & 14.283 & $ -0.101 $ & 61 & 59 & 0.044 & 0.039 & Ell & --\\
BRIGHT-LMC-ECL-17 & 5\uph34\upm41\zdot\ups4 & $ -69\arcd31\arcm39\arcs $ & LMC168.1 &    1.404800 & 13.885 & 13.850 & $ -0.036 $ & 37 & 41 & 0.129 & 0.128 & EW & MACHO 81.8763.8\\
BRIGHT-LMC-ECL-18 & 5\uph06\upm00\zdot\ups3 & $ -70\arcd40\arcm54\arcs $ & LMC121.1 &    1.668340 & 13.614 & 15.306 & $  1.692 $ & 23 &  5 & 0.644 & 0.320 & EW & --\\
BRIGHT-LMC-ECL-19 & 5\uph05\upm11\zdot\ups9 & $ -70\arcd11\arcm20\arcs $ & LMC121.5 &    1.698480 & 14.213 & 13.963 & $ -0.250 $ & 57 & 46 & 0.032 & 0.031 & Ell & --\\
BRIGHT-LMC-ECL-20 & 4\uph49\upm51\zdot\ups4 & $ -69\arcd12\arcm04\arcs $ & LMC135.6 &    1.736348 & 12.596 & 12.541 & $ -0.056 $ & 63 & 66 & 0.030 & 0.031 & EW & 2MASS J04495141-6912043\\
BRIGHT-LMC-ECL-21 & 5\uph37\upm59\zdot\ups5 & $ -69\arcd09\arcm02\arcs $ & LMC175.6 &    1.855280 & 13.711 & 13.744 & $  0.033 $ & 59 & 56 & 0.029 & 0.020 & Ell & 2MASS J05375943-6909010\\
BRIGHT-LMC-ECL-22 & 5\uph40\upm20\zdot\ups6 & $ -69\arcd39\arcm01\arcs $ & LMC176.4 &    1.871260 & 14.477 & 14.461 & $ -0.016 $ & 58 & 58 & 0.044 & 0.042 & EW & [FBM2009] 107\\
BRIGHT-LMC-ECL-23 & 5\uph28\upm60\zdot\ups0 & $ -68\arcd49\arcm27\arcs $ & LMC167.8 &    1.902040 & 14.332 & 14.182 & $ -0.150 $ & 41 & 17 & 0.041 & 0.037 & Ell & SSTISAGEMC J052859.98-684926.5\\
BRIGHT-LMC-ECL-24 & 4\uph54\upm27\zdot\ups5 & $ -68\arcd33\arcm52\arcs $ & LMC134.3 &    2.001460 & 12.700 & 13.470 & $  0.770 $ & 78 & 69 & 0.014 & 0.019 & Ell & --\\
BRIGHT-LMC-ECL-25 & 5\uph06\upm29\zdot\ups0 & $ -70\arcd37\arcm51\arcs $ & LMC121.1 &    2.176380 & 14.271 & 14.058 & $ -0.213 $ & 63 & 57 & 0.072 & 0.071 & EW & --\\
\noalign{\vskip3pt}
\hline
}\end{landscape}
\begin{figure}[htb]
\centerline{\includegraphics[width=13.3cm]{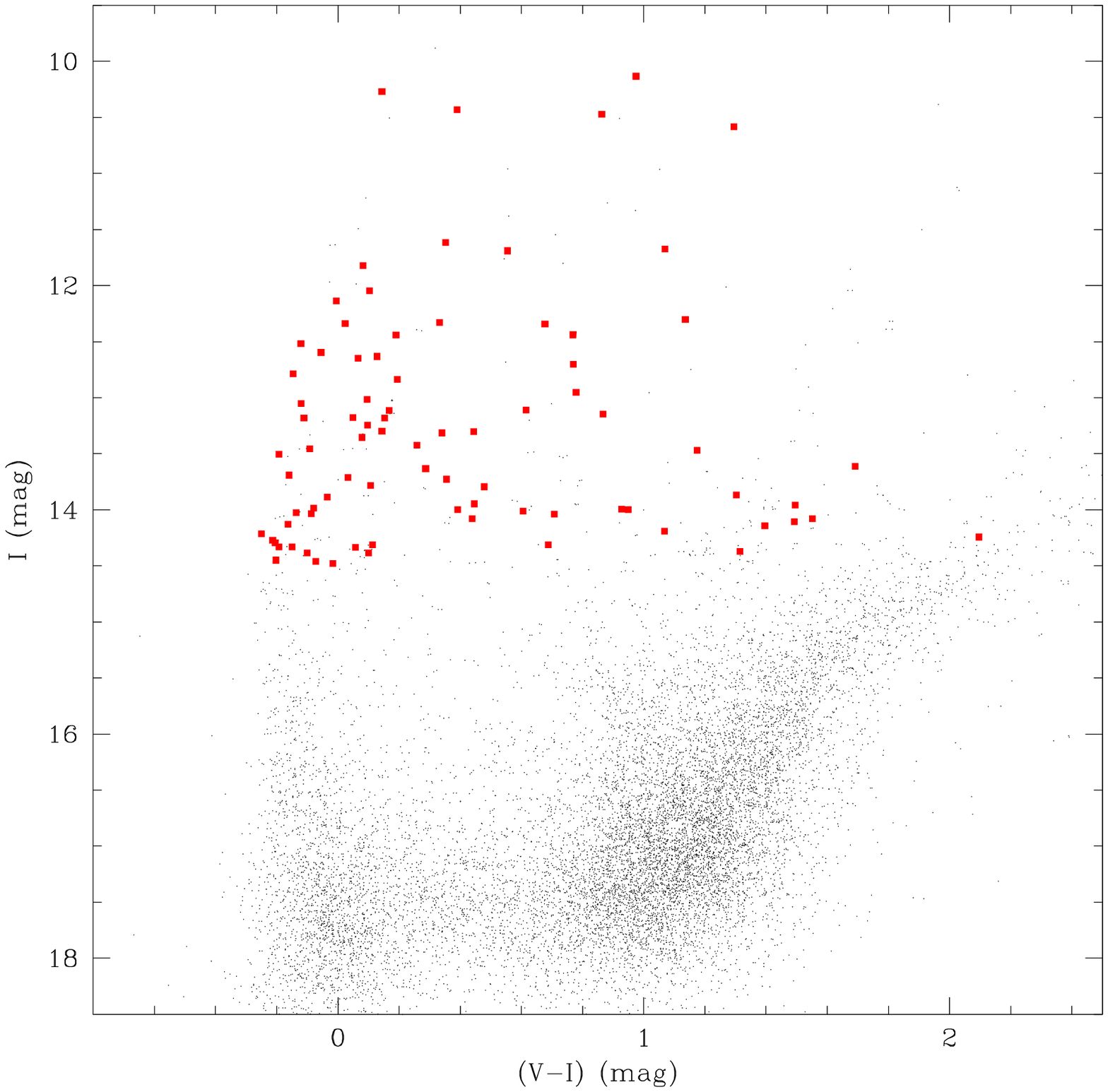}}
\FigCap{Color--magnitude diagram for eclipsing stars (red) added to the
OGLE-III catalog from Shallow Survey data. Black dots represent all stars
from the LMC100.1 subfield.}
\end{figure}

\Section{Miras and Semiregular Variables}
Amongst red giants eight new Miras and 102 semiregular variables (SRV) were
found. They belong to the broad category of Long-Period Variables (LPV).
The cross-identification with other catalogs shows that three classified
Miras were already found while four other objects were recognized as
variable stars. Compared to the main OGLE-III catalog (consisting of
12\,795 objects in the full magnitudes range) this sample is very small so
it cannot affect noticeably already conducted analysis (Soszyñski \etal
2009b), especially as stars of those types split into several groups with
different luminosity--period relations. Where possible each star was
cross-identified with 2MASS Survey infrared counterparts (Cutri \etal
2003). In Fig.~10 color--magnitude diagram is presented with cyan points
representing SRV stars and red points representing Miras. Part of the
catalog with basic data for those objects is presented in Table~3. The
columns contain following data: (1)~variable star identification label,
\begin{figure}[htb]
\centerline{\includegraphics[width=14cm, clip,trim=0 10 0 20]{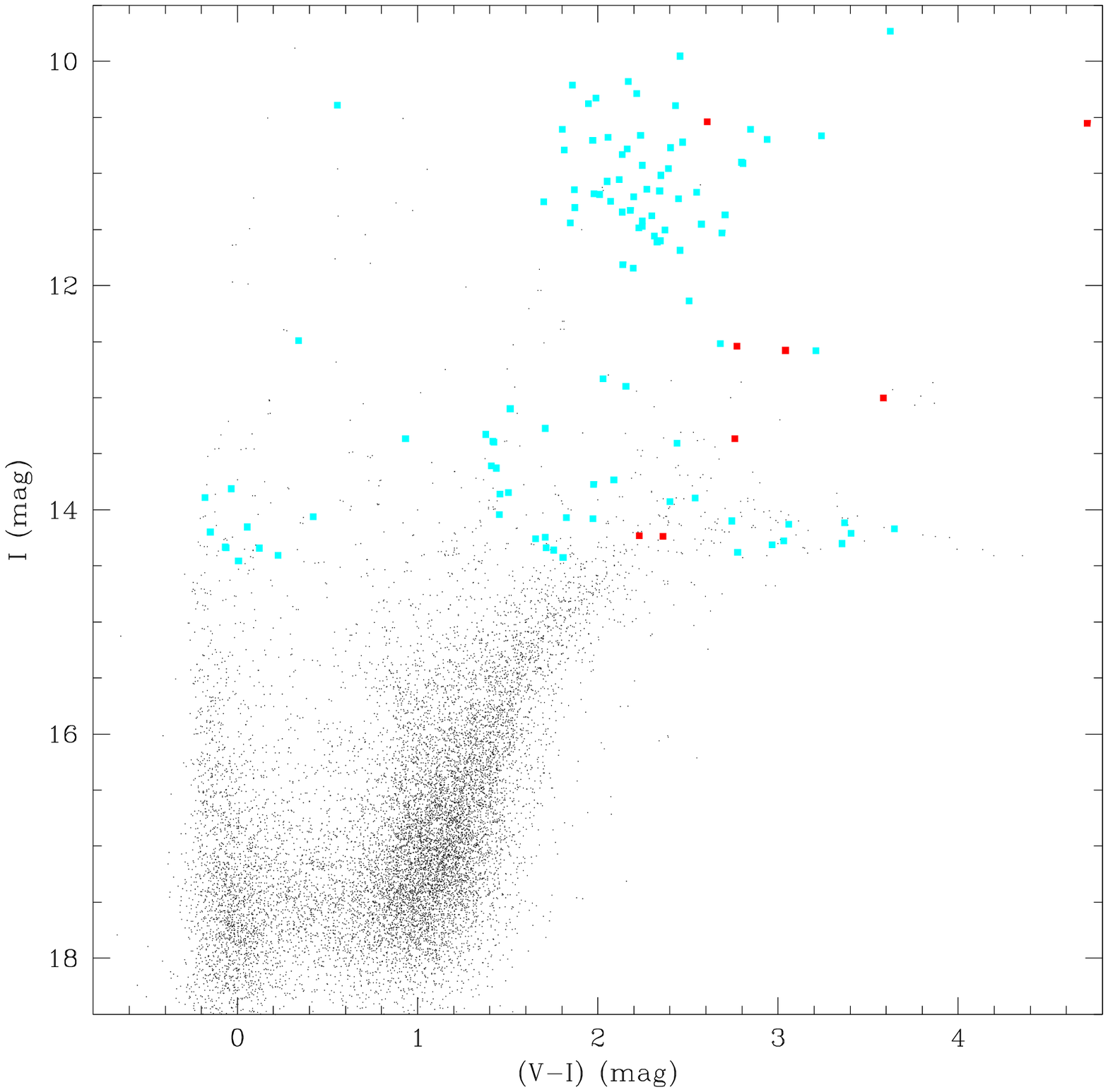}}
\FigCap{Color--magnitude diagram for long-period variables. Red points 
represent Miras and cyan points correspond to SRV stars. Black dots represent all stars
from the LMC100.1 subfield.}
\end{figure}
(2,3)~equatorial coordinates J2000.0, (4)~subfield designation,
(5)~dominant period, (6)~{\it I}-band mean magnitude, (7)~{\it V}-band mean
magnitude, (8)~({\it V-I}) color index, (9,10)~number of observations in
the {\it I}- and {\it V}-band, (11,12)~dispersion of magnitudes for the
{\it I}- and {\it V}-band, (13)~subtype, (14-17)~2MASS object
cross-identification: separation distance and {\it JKH} infrared
photometry, (18)~designations in other catalogs/surveys (the same as for
eclipsing variables). Fig.~11 shows {\it I}-band light curves for new
Miras.

\begin{landscape}
\renewcommand{\TableFont}{\tiny}
\renewcommand{\arraystretch}{1.25}
\MakeTableSep{ccccccccccccc|@{\hspace{2pt}}c@{\hspace{4pt}}c@{\hspace{4pt}}c@{\hspace{4pt}}c|c}
{22cm}{First 25 lines of long-period variable stars catalog}{
\hline
\uprule
ID & RA & DEC & Subfield & \multicolumn{1}{c}{Period} & {\it I} & {\it V}  & $V-I$ & $N_I$ & $N_V$ & $\sigma_I$ & $\sigma_V$ & Subtype&d&J&H&K&Other\\
\dorule
&(J2000.0) &(J2000.0) & &\multicolumn{1}{c}{[days]} & [mag] & [mag] & [mag]  & & & [mag]&[mag] & &[arcsec] & [mag] & [mag] & [mag]  &designations\\
\hline
\uprule
BRIGHT-LMC-LPV-001 & 5\uph24\upm04\zdot\ups7 & $ -70\arcd33\arcm10\arcs $ & LMC163.7 &  178.571430 & 14.233 & 16.464 &  2.231& 38 & 19 & 0.390 & 0.719 & MIRA & 1.35 & 12.992 & 12.189 & 11.783 & --\\
BRIGHT-LMC-LPV-002 & 5\uph25\upm11\zdot\ups6 & $ -68\arcd42\arcm44\arcs $ & LMC160.7 &  248.138960 & 12.578 & 15.619 &  3.041& 42 & 17 & 0.390 & 0.606 & MIRA & 0.40 & 10.307 &  9.395 &  9.026 & 2MASS J05251160-6842444\\
BRIGHT-LMC-LPV-003 & 4\uph36\upm46\zdot\ups8 & $ -70\arcd18\arcm41\arcs $ & LMC151.6 &  250.000000 & 10.552 & 15.268 &  4.716& 25 & 22 & 0.777 & 1.823 & MIRA & 0.04 &  7.729 &  6.917 &  6.500 & 043648-7018.6\\
BRIGHT-LMC-LPV-004 & 5\uph15\upm47\zdot\ups4 & $ -70\arcd04\arcm33\arcs $ & LMC112.1 &  261.780100 & 14.236 & 16.598 &  2.362& 63 & 24 & 0.713 & 0.727 & MIRA & 1.47 & 12.442 & 11.595 & 11.193 & --\\
BRIGHT-LMC-LPV-005 & 5\uph39\upm56\zdot\ups9 & $ -69\arcd35\arcm21\arcs $ & LMC176.4 &  416.739126 & 12.541 & 15.312 &  2.770& 59 & 52 & 0.413 & 0.890 & MIRA & 0.29 & 10.759 &  9.900 &  9.589 & HV 2763\\
BRIGHT-LMC-LPV-006 & 5\uph27\upm10\zdot\ups2 & $ -69\arcd36\arcm27\arcs $ & LMC162.4 &  512.170000 & 13.004 & 16.587 &  3.583& 55 & 19 & 0.881 & 1.026 & MIRA & 0.31 & 10.399 &  9.544 &  9.159 & HV 12048\\
BRIGHT-LMC-LPV-007 & 5\uph40\upm32\zdot\ups8 & $ -71\arcd31\arcm59\arcs $ & LMC179.6 &  625.000000 & 13.367 & 16.128 &  2.762& 52 & 11 & 0.745 & 1.041 & MIRA & 0.15 & 10.224 &  9.349 &  8.829 & 2MASS J05403279-7131591\\
BRIGHT-LMC-LPV-008 & 5\uph30\upm41\zdot\ups4 & $ -69\arcd15\arcm34\arcs $ & LMC168.7 &  669.230782 & 10.540 & 13.147 &  2.607& 38 & 44 & 0.412 & 0.805 & MIRA & 0.31 &  8.752 &  7.993 &  7.591 & 053041-6915.5\\
BRIGHT-LMC-LPV-009 & 5\uph08\upm21\zdot\ups3 & $ -67\arcd55\arcm18\arcs $ & LMC117.3 &   75.757580 & 13.735 & 15.824 &  2.089& 21 & 17 & 0.043 & 0.087 & SRV & 1.40 & 12.161 & 11.207 & 10.925 & --\\
BRIGHT-LMC-LPV-010 & 5\uph31\upm22\zdot\ups8 & $ -69\arcd02\arcm30\arcs $ & LMC168.5 &   82.987550 & 14.457 & 14.463 &  0.006& 57 & 56 & 0.046 & 0.035 & SRV & 0.28 & 14.407 & 14.558 & 14.594 & MACHO 82.8286.14\\
BRIGHT-LMC-LPV-011 & 5\uph05\upm29\zdot\ups7 & $ -69\arcd00\arcm27\arcs $ & LMC119.5 &   97.087380 & 13.926 & 16.326 &  2.400& 63 & 58 & 0.090 & 0.196 & SRV & -- & -- & -- & -- & --\\
BRIGHT-LMC-LPV-012 & 5\uph15\upm41\zdot\ups0 & $ -69\arcd01\arcm10\arcs $ & LMC111.4 &  107.526880 & 14.042 & 15.495 &  1.453& 73 & 42 & 0.033 & 0.036 & SRV & 0.41 & 13.051 & 12.342 & 12.202 & 2MASS J05154103-6901096\\
BRIGHT-LMC-LPV-013 & 5\uph30\upm18\zdot\ups5 & $ -70\arcd26\arcm41\arcs $ & LMC170.7 &  107.526880 & 14.338 & 16.051 &  1.714& 52 & 23 & 0.031 & 0.059 & SRV & 0.20 & 13.160 & 12.389 & 12.129 & 2MASS J05301846-7026411\\
BRIGHT-LMC-LPV-014 & 5\uph30\upm10\zdot\ups3 & $ -69\arcd00\arcm46\arcs $ & LMC168.5 &  109.890110 & 14.260 & 15.914 &  1.654& 38 & 19 & 0.041 & 0.063 & SRV & 0.04 & 13.079 & 12.300 & 12.094 & 2MASS J05301027-6900460\\
BRIGHT-LMC-LPV-015 & 5\uph29\upm46\zdot\ups2 & $ -68\arcd37\arcm03\arcs $ & LMC167.6 &  113.895220 & 11.328 & 13.509 &  2.181& 30 & 39 & 0.047 & 0.088 & SRV & 0.56 &  9.897 &  9.034 &  8.747 & 2MASS J05294618-6837024\\
BRIGHT-LMC-LPV-016 & 4\uph53\upm44\zdot\ups4 & $ -69\arcd22\arcm54\arcs $ & LMC135.2 &  114.547540 & 13.844 & 15.347 &  1.502& 72 & 60 & 0.221 & 0.034 & SRV & 0.45 & 12.692 & 11.969 & 11.811 & 2MASS J04534436-6922535\\
BRIGHT-LMC-LPV-017 & 5\uph10\upm32\zdot\ups6 & $ -69\arcd50\arcm44\arcs $ & LMC112.7 &  121.951220 & 14.245 & 15.952 &  1.707& 26 & 13 & 0.311 & 0.241 & SRV & -- & -- & -- & -- & --\\
BRIGHT-LMC-LPV-018 & 4\uph58\upm04\zdot\ups0 & $ -71\arcd09\arcm11\arcs $ & LMC130.2 &  131.578950 & 12.516 & 15.196 &  2.679& 41 & 25 & 0.049 & 0.100 & SRV & 0.18 & 10.978 & 10.007 &  9.748 & 2MASS J04580400-7109108\\
BRIGHT-LMC-LPV-019 & 5\uph40\upm17\zdot\ups5 & $ -69\arcd30\arcm57\arcs $ & LMC175.1 &  138.312590 & 10.390 & 10.944 &  0.554& 58 & 50 & 0.028 & 0.033 & SRV & 0.46 &  9.902 &  9.721 &  9.581 & 2MASS J05401751-6930574\\
BRIGHT-LMC-LPV-020 & 5\uph23\upm48\zdot\ups0 & $ -70\arcd07\arcm35\arcs $ & LMC162.8 &  140.056020 & 11.814 & 13.953 &  2.138& 41 & 30 & 0.033 & 0.078 & SRV & 0.36 & 10.407 &  9.545 &  9.269 & 2MASS J05234794-7007348\\
BRIGHT-LMC-LPV-021 & 4\uph49\upm46\zdot\ups8 & $ -69\arcd23\arcm14\arcs $ & LMC135.7 &  144.717800 & 11.844 & 14.041 &  2.197& 74 & 70 & 0.039 & 0.117 & SRV & 0.17 & 10.364 &  9.470 &  9.159 & --\\
BRIGHT-LMC-LPV-022 & 5\uph32\upm58\zdot\ups3 & $ -69\arcd55\arcm51\arcs $ & LMC169.2 &  155.521000 & 13.406 & 15.846 &  2.440& 56 & 52 & 0.052 & 0.092 & SRV & 0.64 & 11.540 & 10.539 & 10.203 & 2MASS J05325832-6955503\\
BRIGHT-LMC-LPV-023 & 5\uph37\upm05\zdot\ups9 & $ -69\arcd52\arcm00\arcs $ & LMC176.7 &  161.290320 & 14.078 & 16.051 &  1.972& 30 & 27 & 0.064 & 0.111 & SRV & 0.50 & 12.739 & 11.842 & 11.583 & 2MASS J05370582-6951598\\
BRIGHT-LMC-LPV-024 & 5\uph37\upm20\zdot\ups5 & $ -69\arcd19\arcm39\arcs $ & LMC175.7 &  162.601630 & 10.704 & 12.676 &  1.972& 59 & 56 & 0.032 & 0.063 & SRV & 0.36 &  9.418 &  8.541 &  8.278 & 2MASS J05372049-6919386\\
\dorule
BRIGHT-LMC-LPV-025 & 5\uph08\upm59\zdot\ups7 & $ -68\arcd35\arcm36\arcs $ & LMC118.3 &  164.744650 & 14.423 & 16.230 &  1.807& 73 & 66 & 0.029 & 0.055 & SRV & 0.69 & 13.198 & 12.294 & 12.079 & --\\
\hline
}
\end{landscape}
\begin{figure}[htb]
\vglue-10mm
\centerline{\includegraphics[width=13cm]{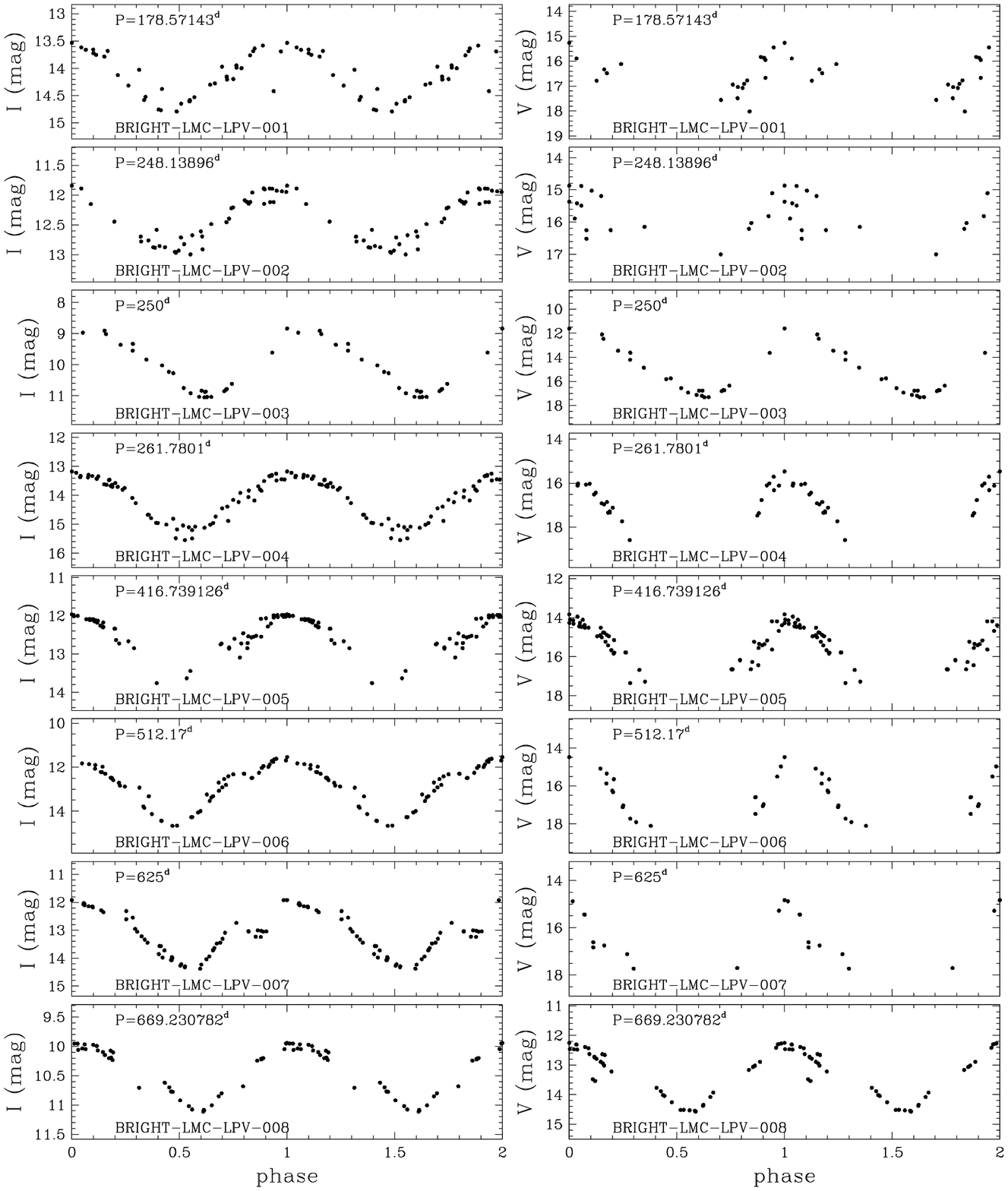}}
\vspace*{-11mm}
\FigCap{{\it VI}-band light curves for new Miras.}
\end{figure}
\Section{Miscellaneous Variables}
The most numerous group of variable stars in our sample constitute stars
for which we could not clearly determine periodicity. We found 558 such
objects. Due to limited observational epochs it is possible we were not
able to distinguish some, in fact, periodic variable stars. Long-period
variables often require observations in long timescales to establish the
nature of their brightness fluctuations. This condition is fulfilled by
OGLE project data (now it is over 20 years of regular survey) but
unfortunately Shallow Survey spanned for 4.5 years only and for some fields
only 13 epochs were collected. However in this way we show which stars can
be particularly interesting for further follow-up observations. In order to
have a broader view of our sample we prepared color--magnitude diagram
(Fig.~12) with color-coded ``amplitudes'' (differences between maximum and
minimum magnitudes after removal of $3\sigma$ deviating points). One can
notice that bigger-amplitude variables concentrate in two regions of the
CMD. We also checked that variables do not separate on Wesenheit indices
$W_I$--$W_{JK}$ plane (Fig.~13), what takes place in case of long-period
variables.
\begin{figure}[h]
\vspace*{-6mm}
\centerline{\includegraphics[width=13.5cm]{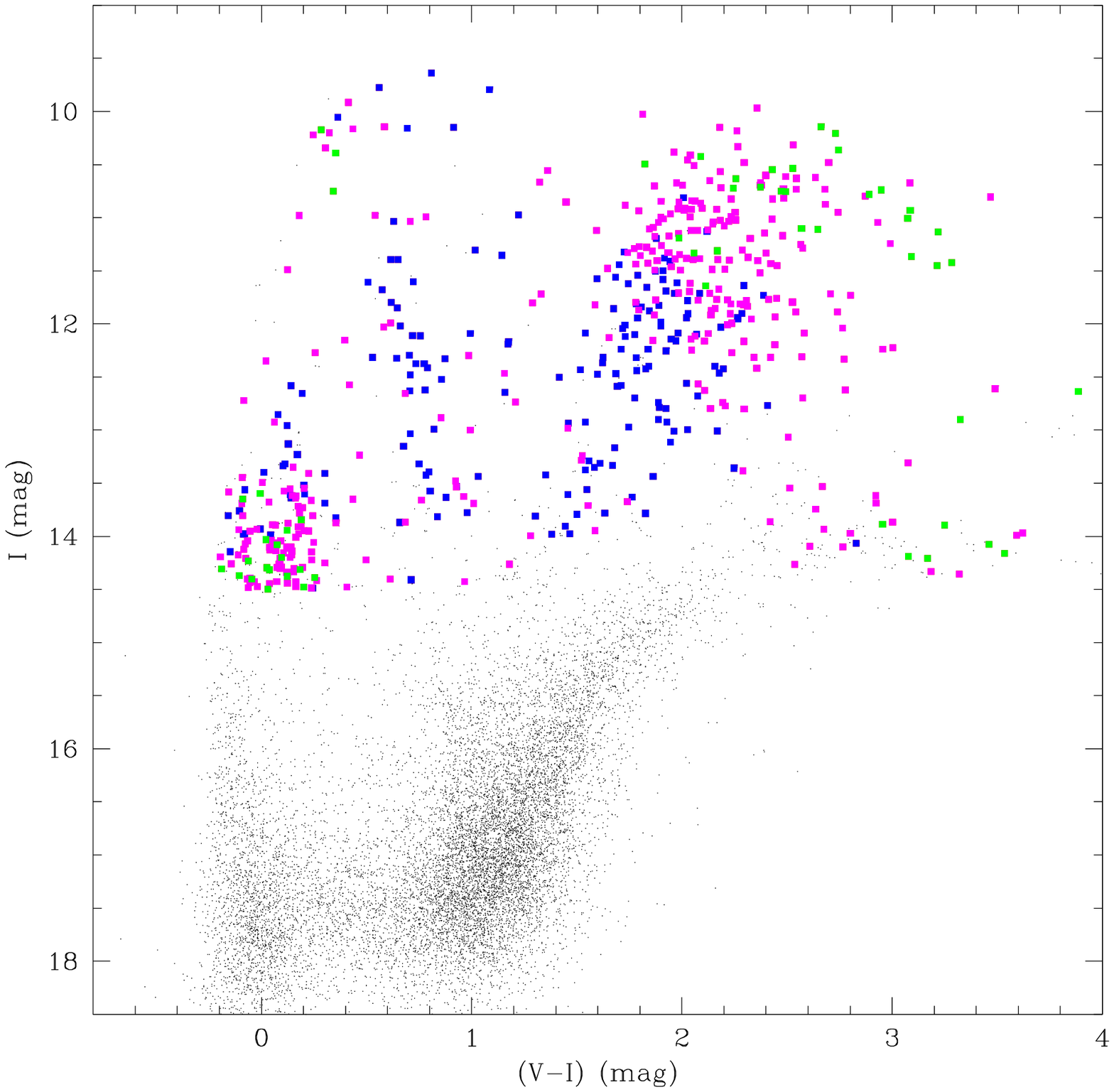}}
\vspace*{-4mm}
\FigCap{Color--magnitudes diagram for miscellaneous variables. Amplitudes
are color-coded: blue -- ${\rm amp\le0.1}$~mag, magenta -- ${\rm 0.1~mag<amp\le0.5}$
mag, green -- ${\rm amp>0.5}$~mag. Black dots represent all stars
from the LMC100.1 subfield.}
\end{figure}
\begin{figure}[htb]
\centerline{\includegraphics[width=13cm]{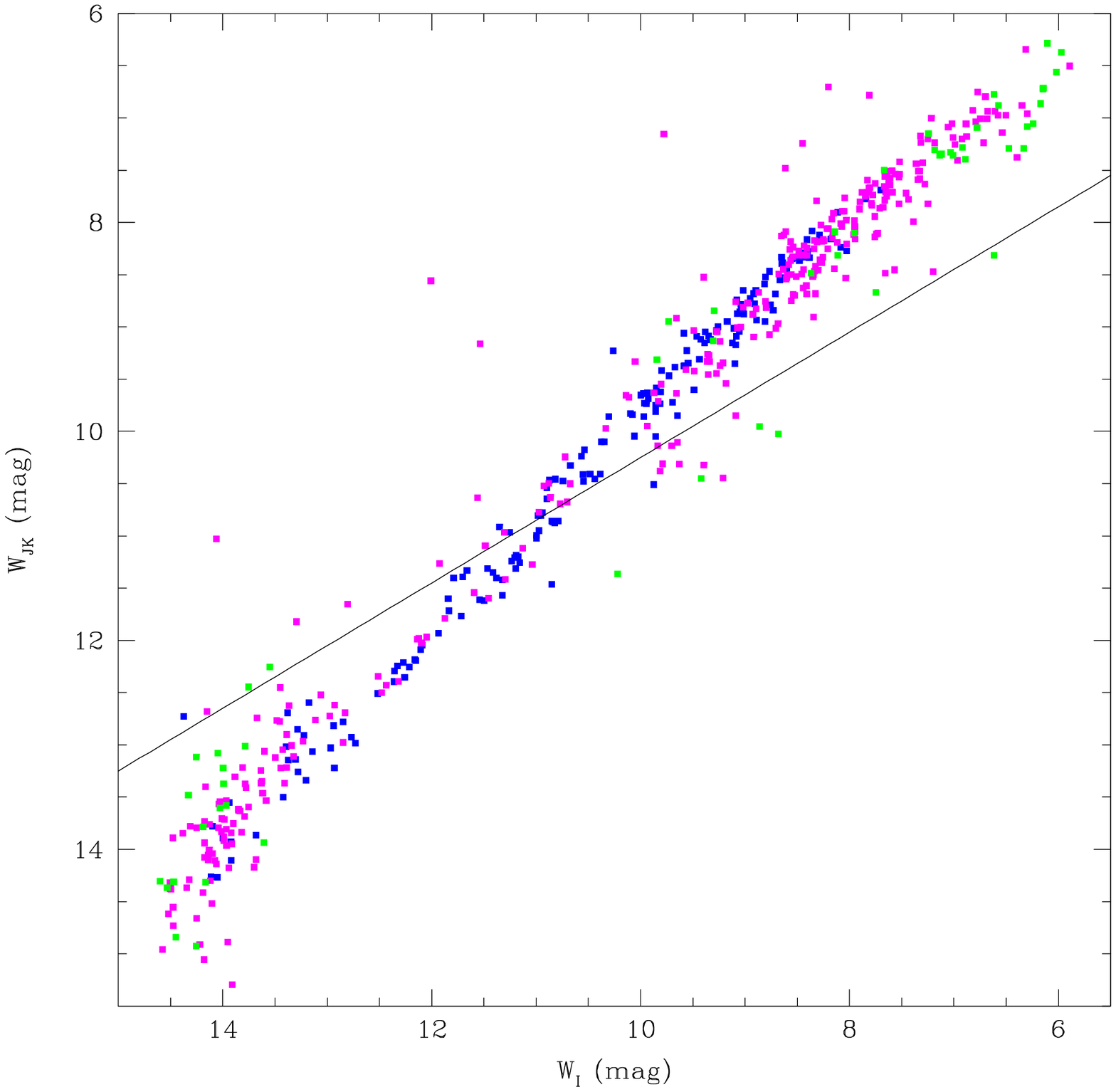}}
\FigCap{Wesenheit indices diagram for miscellaneous variables 
cross-identified with 2MASS catalog. $W_{JK}$ denotes infrared Wesenheit
index defined as $W_{JK}=J-0.686\cdot(J-K)$. Amplitudes are color-coded: blue
-- ${\rm amp\le0.1}$~mag, magenta -- ${\rm 0.1~mag<amp\le0.5}$~mag, green
-- ${\rm amp>0.5}$~mag. Black line represents approximate separation border
between oxygen-rich and carbon-rich stars based on data from Soszyñski
\etal (2009b).}
\end{figure}

Basic parameters for those stars are presented in OGLE Internet Archive in
the same manner as for long-period variables with the ``period'' column
omitted.

\Section{Discussion}
The data from the Shallow Survey allowed us to extend period--luminosity
relation for classical Cepheids in the Large Magellanic Cloud up to 134
days. In that range we did not observe any deviations from linearity.
However, this conclusion is a little weakened due to the presence of only
two Cepheids with periods above 53~days in our sample. Therefore we cannot
confirm recent suggestion that for long-period Cepheids ($P>100$~d)
luminosity is period--independent (Bird \etal 2009) neither lack of such a
flattening (Fiorentino \etal 2012). We have also updated long-period
variables and eclipsing binaries catalogs with new objects of luminosities
up to 9.7~mag. The selected sample of miscellaneous stars of suspected
irregular variability is worth performing follow-up observations. It is
still an open question if red giants can pulsate in strictly irregular
manner.

The fact that we have not found significant number of new regular variable
stars proves that the OGLE-III catalogs are very complete. Variable stars
were omitted mainly in specific situations -- like a saturation of the star
itself or its location near other saturated object.

\vspace*{-7pt}
\Section{Data Availability}
\vspace*{-4pt}
The described supplementary catalogs of bright variable stars in the LMC
and their photometry are available to the astronomical community from the
OGLE Internet Archive:

\centerline{\it http://ogle.astrouw.edu.pl}

\centerline{\it ftp://ftp.astrouw.edu.pl/ogle/ogle3/OIII-CVS/bright\_lmc/}

\Acknow{The OGLE project has received funding from the European Research 
Council under the European Community's Seventh Framework Programme
(FP7/2007-2013)/ERC grant agreement no. 246678 to AU. This paper was
partially supported by Polish grant N~N203 510038. We gratefully
acknowledge financial support for this work from the Chilean Center for
Astrophysics FONDAP 15010003, and from the BASAL Centro de Astrofisica y
Tecnologias Afines (CATA) PFB-06/2007.}

\end{document}